\documentclass[fleqn,usenatbib]{mnras}

\usepackage{newtxtext,newtxmath}
\usepackage{soulutf8}
\usepackage{xcolor}
\sethlcolor{yellow}

\usepackage[T1]{fontenc}

\DeclareRobustCommand{\VAN}[3]{#2}
\let\VANthebibliography\thebibliography
\def\thebibliography{\DeclareRobustCommand{\VAN}[3]{##3}\VANthebibliography}

\usepackage{graphicx}
\usepackage{caption,subcaption}
\usepackage{geometry}

\title[Edge-on galaxies with MUSE --I. ESO 544-27]{Understanding stellar populations in thin \& thick discs of edge-on galaxies with MUSE -- I. The case of the reignited S0 galaxy ESO~544-27}

\author[Somawanshi, Bhattacharya et al.]
{Devang Somawanshi,$^{1}$\thanks{E-mail: somawanshi20@iiserb.ac.in}
Souradeep Bhattacharya,$^{2}$\thanks{E-mail: souradeep@iucaa.in}
Manish Kataria,$^{2}$ and
Chiaki Kobayashi,$^{3}$\\
{$^{1}$Department of Physics, Indian Institute of Science Education and Research Bhopal, Bhopal Bypass Road, Bhauri, Bhopal 462 066, Madhya Pradesh, India}\\
{$^{2}$Inter University Centre for Astronomy and Astrophysics, Ganeshkhind, Post Bag 4, Pune 411007, India}\\
{$^{3}$Centre for Astrophysics Research, Department of Physics, Astronomy and Mathematics, University of Hertfordshire, Hatfield, AL10 9AB, UK}
}

\date{Accepted XXX. Received YYY; in original form ZZZ}

\pubyear{2023}

\begin{document}
\label{firstpage}
\pagerange{\pageref{firstpage}--\pageref{lastpage}}
\maketitle

\begin{abstract}
Edge-on galaxies act as the best laboratories to understand the origin of thin and thick discs in galaxies. Measurement of spatially resolved stellar population properties in such galaxies, particularly age, metallicity and [$\alpha$/Fe], are crucial to understanding the formation and evolution of disc galaxies. Such measurements are
made possible from stellar population model fits to deep integral field spectroscopic (IFU) observations of resolved galaxies. We utilise archival MUSE IFU observations of the edge-on galaxy ESO~544-27 to uncover the formation history of its thin and thick discs through its stellar populations. We find the thin disc of the galaxy is dominated by an old ($>9$~Gyr) low [$\alpha$/Fe] metal-rich stellar population. Its outer thick disc is dominated by an old ($>9$~Gyr) high [$\alpha$/Fe] metal-rich component that should have formed with higher star-formation efficiency than the Milky Way thick disc. We thus find [$\alpha$/Fe] dichotomy in ESO~544-27 with its thin and thick discs dominated by low and high [$\alpha$/Fe] stellar populations respectively. However, we also find a metal-rich younger ($<2$~Gyr old) stellar population in ESO~544-27. The galaxy was nearly quenched until its star-formation was reignited recently first in the outer and inner thick disc ($\sim$1~Gyr ago) and then in the thin disc ($\sim$600~Myr ago). We thus find that both the low [$\alpha$/Fe] thin and high [$\alpha$/Fe] thick discs of ESO~544-27 are inhabited primarily by similarly old metal-rich stellar populations, a contrast to that of other galaxies with known thin and thick disc stellar population properties.
\end{abstract}

\begin{keywords}
galaxies: individual: ESO 544-27 -- galaxies: abundances -- galaxies: evolution -- galaxies: formation -- galaxies: ISM -- galaxies: spiral -- galaxies: stellar content -- galaxies: structure
\end{keywords}


\section{Introduction}
\label{sect: intro}

Thick discs in galaxies have been defined as rotating components with large-scale height and high-velocity dispersion having an exponentially-decaying surface brightness profile \citep[see review by][and references therein]{vanderKruit11}. The term `thick disc' was introduced by \citet {BURSTEIN1979} to describe a newly identified diffuse region of stars in edge-on S0 galaxies. 
In the Milky Way galaxy, the thick disc was discovered and described by \citet {GilmoreReed1983} with a relatively high scale height compared to the thin disc, having old and metal-poor stars. Subsequent observations of other edge-on galaxies revealed the ubiquity of photometrically distinct thin and thick discs  \citep{2002Dalcantonbernstein, Comerón2018}. 
 

The formation and evolution of thick discs in galaxies have been the subject of many studies \citep{vanderKruit11} but the stellar population of these discs have been studied in most detail in the Milky Way (MW), with the stars in the thick disc being mostly older than those in the thin disc \citep[see review by][and references therein]{2016BHawthorne}. From chemical abundances determined from individual stars, the differences between the stellar populations in the two discs have been identified. The thick disc is mostly dominated by older high [$\alpha$/Fe] (more enriched in $\alpha$ elements) \& metal-poor stars while the thin disc is dominated by younger low [$\alpha$/Fe] (less enriched in $\alpha$ elements) \& metal-rich stars \citep[e.g.][]{1998Fuhrmann}. Properties of the thin and thick disc stars in the [$\alpha$/Fe] vs [Fe/H] plane depends on the position of the examined stars in the galaxy \citep[][]{2015Hayden,Guiglion23}.  Such dependence is also seen in chemodynamical simulations \citep{Kobayashi16,Vincenzo20}. The stars in the MW thick disc are thought to have formed relatively rapidly at early times thereby forming a more high [$\alpha$/Fe] \& metal-poor old population ($\geq10$~Gyr) while those in the MW thin disc are thought to have formed over a more extended period of star formation (since $\sim$10~Gyr ago) that continues to the present day \citep{2016BHawthorne}. External gas-infall can also leave imprints onto the [$\alpha$/Fe] vs [Fe/H] plane, as has been claimed for the MW \citep{Spitoni19}.

  
In the Andromeda galaxy (M~31), the nearest giant spiral galaxy to the MW \citep[$776.2^{+22}_{-21}$~kpc;][]{Savino22} and the most massive member of our local group at $\sim 1.5 \times 10^{12}$~M$_{\odot}$ \citep[see][and references therein]{Bhattacharya23b}, the kinematically and chemically distinct thin and thick discs of M31 were identified \citep{Bh+19b,Bhattacharya22} from the survey of planetary nebulae in M~31 \citep{Bh+19,Bh21}. Chemical abundances of M~31 disc planetary nebulae, in conjunction with galactic chemical evolution models \citep{Arnaboldi22,Kobayashi23}, indicate that the older, high [$\alpha$/Fe] thick-disc population of M~31 formed through a more intense initial starburst than in the Milky Way. In contrast, the M~31 thin disc resulted from a burst 2.5–4.5 billion years ago, involving relatively metal-poor gas mixed with pre-enriched gas within the M 31 disc. This is consistent with a wet major-merger (mass ratio $\sim$ 1:4) scenario in M~31 \citep{Hammer18, Bhattacharya23a}.

While the origin of the MW and M~31 thick and thin discs have some constraints from stellar population properties from the aforementioned studies, similar studies in other galaxies have been limited. Using long-slit spectroscopic observations, \citet{2008Yoachim} measured Lick index equivalent widths \citep{Burstein84,Faber85} to derive luminosity weighted stellar ages and metallicities for thin and thick-discs of nine edge-on disc galaxies. They found that all their galaxies had old (4-10~Gyr old) thick discs and relatively younger thin discs. However, they could not determine substantial metallicity differences between the thin and thick discs for the galaxies in their sample.

Using stellar population model fits to long-slit spectra of three edge-on S0 galaxies (taken at the mid-plane for thin disc and at large scale-height for thick disc), \citet{2016Kasparova} found diversity in disc stellar populations properties. NGC~4111 (distance$\sim$15~Mpc) showed an intermediate-age ($\sim$5~Gyr), metal-rich ([Fe/H]$\sim$-0.2), low [$\alpha$/Fe] thick disc and similar aged but slightly more metal-poor ([Fe/H]$\sim$-0.4) low [$\alpha$/Fe] thin disc. NGC~4710 (distance$\sim$16.5~Mpc) also showed an intermediate-age ($\sim$5~Gyr), metal-rich ([Fe/H]$\sim$-0.2) low [$\alpha$/Fe]  thick disc though the thin disc was younger ($\sim$2.5~Gyr) but slightly more metal-rich ([Fe/H]$\sim$0.0) but also low [$\alpha$/Fe]. NGC~5422 (distance$\sim$30.9~Mpc) showed old ($\sim$10~Gyr), metal-rich ([Fe/H]$\sim$-0.2) but slightly higher [$\alpha$/Fe] thick and thin discs. It is unclear if the off-plane measurements truly correspond to the thick discs in these galaxies.

Using stellar population model fits to integrated-field spectroscopy (VIMOS at the ESO VLT; \citealt{LeFevre03}) observations of the edge-on galaxy ESO 533-4, \citet{2015Comeron} identified an old ($\sim10$~Gyr old) metal-poor stellar population in the thick disc, and a relatively younger, more metal-rich stellar population in the thin disc. With similar stellar population model fits to MUSE (at the ESO VLT; \citealt{2010Bacon}) integrated-field spectroscopy observations of the edge-on S0 galaxy ESO 243-49. \citet{Comeron2016} found an old population in both the thin and thick discs, though the former was relatively more metal-rich.

From more MUSE observations of three edge-on S0 galaxies (FCC 170, FCC 153, FCC 177) in the Fornax cluster, \citet[][]{2019PinnaA,2019PinnaB} found that all three galaxies have [$\alpha$/Fe]-rich, metal-poor, old thick discs. FCC 170 also has a significant fraction of younger stars ($\sim10$~Gyr old) characterized by even lower-metallicity and higher [$\alpha$/Fe] values, consistent with being accreted material from disrupted satellites. The outer regions of the thick discs of FCC 170, {FCC 153 and FCC 177 show metal-rich low [$\alpha$/Fe] stars which may be consistent with being material formed in the thin disc of these galaxies that have been dynamically heated to higher scale-heights. FCC 170 has an old but metal-rich low [$\alpha$/Fe] thin disc. FCC 153 and FCC 177 have metal-rich low [$\alpha$/Fe] thin discs that show bimodality in age. In addition to the older thin disc ($\geq8$~Gyr), a population of younger thin disc stars ($\sim$3~Gyr old) also exists, especially in FCC 177 where the younger component has higher mass fraction than the older component}. 

Analyzing UGC 10738, an edge-on MW-like disc galaxy at around 100 Mpc using the same method, Scott et al. (2021) found a stellar population similar to the MW: a high [$\alpha$/Fe], metal-poor, old thick disc, and a relatively low [$\alpha$/Fe], metal-rich, slightly younger thin disc. From a similar study of the more nearby ($\sim26.5$~Mpc) similarly massive edge-on galaxy NGC~5746, \citet{2021MMarie} found similar high [$\alpha$/Fe], metal-poor old thick disc and low [$\alpha$/Fe], metal-rich, younger thin disc. They additionally found that NGC~5746 has experienced a merger with a $\sim$1:10 mass ratio with the disrupted satellite debris contributing almost $\sim30\%$ of the mass of the thick disc. 

To further explore the diversity in disc stellar populations with the goal of eventually uncovering the various thick disc formation scenarios, we explore the MUSE observations of star-forming edge-on galaxies presented by \citet{Comerón2019}. These galaxies were chosen for spectroscopic follow-up from a sample of 70 edge-on galaxies originally part of the S$^{4}$G survey \citep{Comerón2012}. Kinematic properties of the thin and thick discs of a sample of eight galaxies were studied in \citet{Comerón2019}, six of which had distinct thin and thick discs based on photometric decomposition from \citet{Comerón2018}. 

\begin{figure}
\includegraphics[width=\columnwidth]{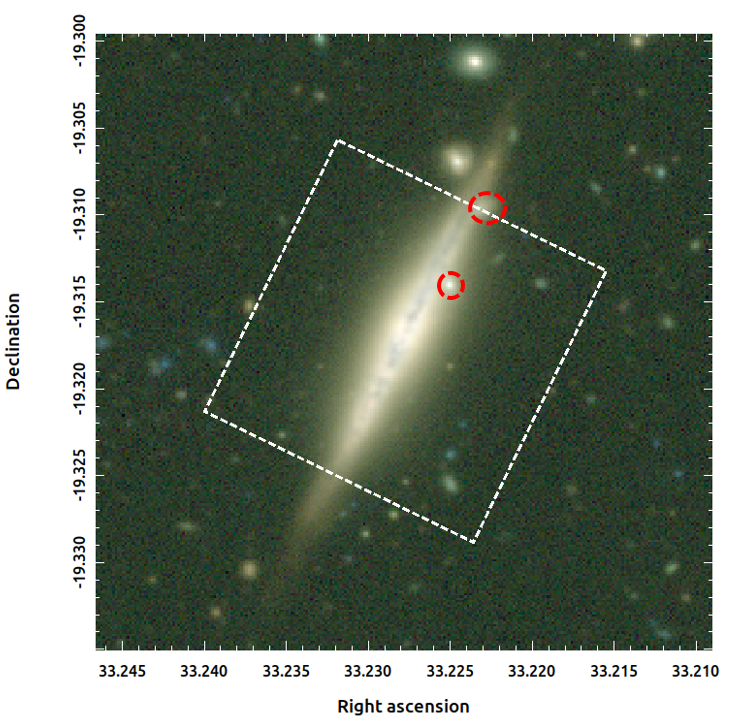}
\caption{Colour image of ESO~544-27 constructed from \textit{g,r,z}-band images from the Dark Energy Camera Legacy Survey (DECaLS DR9, \citealt{Dey19}). The MUSE field-of-view is marked in white. The two interloping galaxies that were removed from the analysed MUSE datacube are marked in red.}
\label{Fig: ESO 544-27}
\end{figure}

In this paper, we carry out a pilot stellar population analysis of ESO~544-27, which is one of the galaxies with photometrically distinct thin and thick discs presented in \citet{Comerón2019}. This study will later be expanded to the other seven edge-on galaxies. We describe the basic properties of ESO~544-27 as well as the MUSE data in Section~\ref{sect: obs}. The stellar population analysis method is described in Section~\ref{sect: analysis}. We derive the galaxy stellar population properties in Section~\ref{Sect: Results}. We discuss the origin of the thin and thick disc of ESO~544-27 in Section~\ref{sect: discussion}. We conclude in Section~\ref{sect: conclusion}.

\begin{figure*}
    \includegraphics[width=\textwidth]{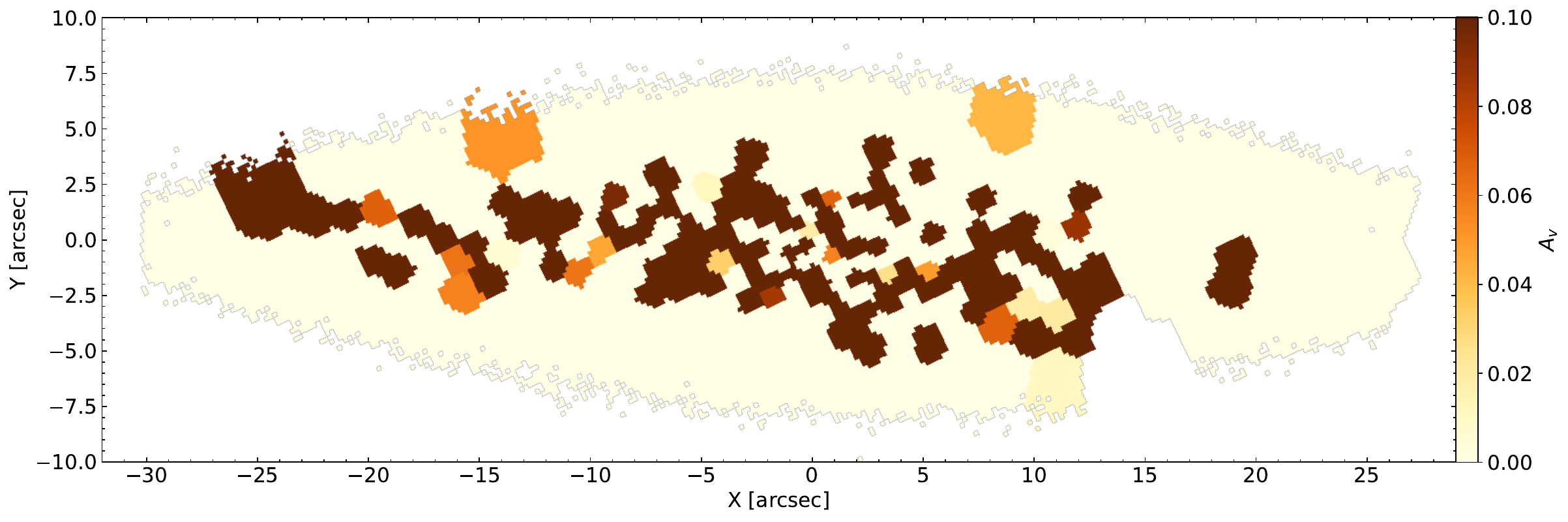}
    

    \caption{Spatial distribution of extinction ($A_{v}$) for each spatial bin in ESO~544-27.}
    \label{Fig: Dust_correction}
\end{figure*}

\begin{figure*}
    \includegraphics[width=\textwidth]{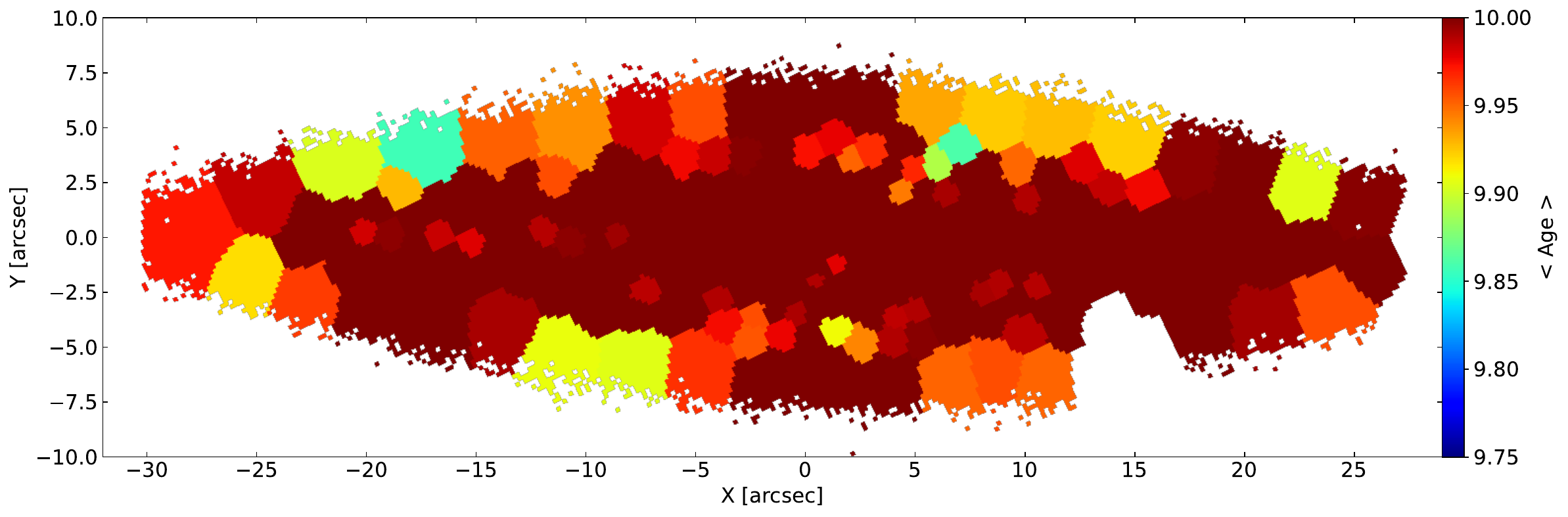}
    
    
    \includegraphics[width=\textwidth]{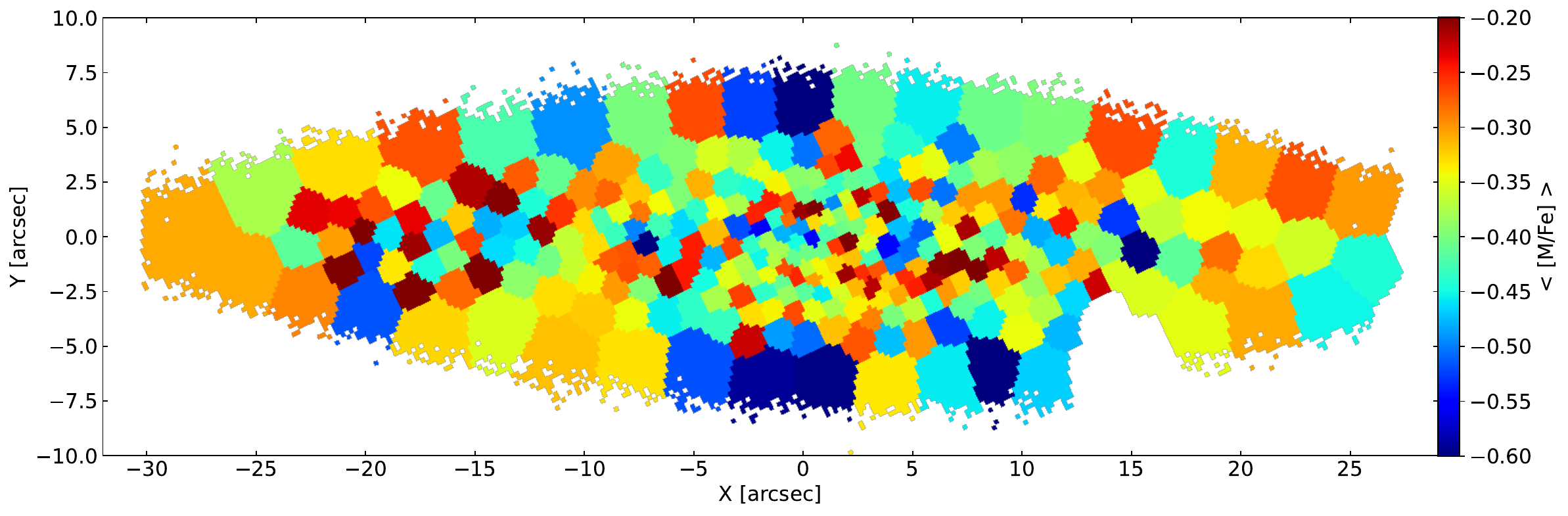}
    
    
    \includegraphics[width=\textwidth]{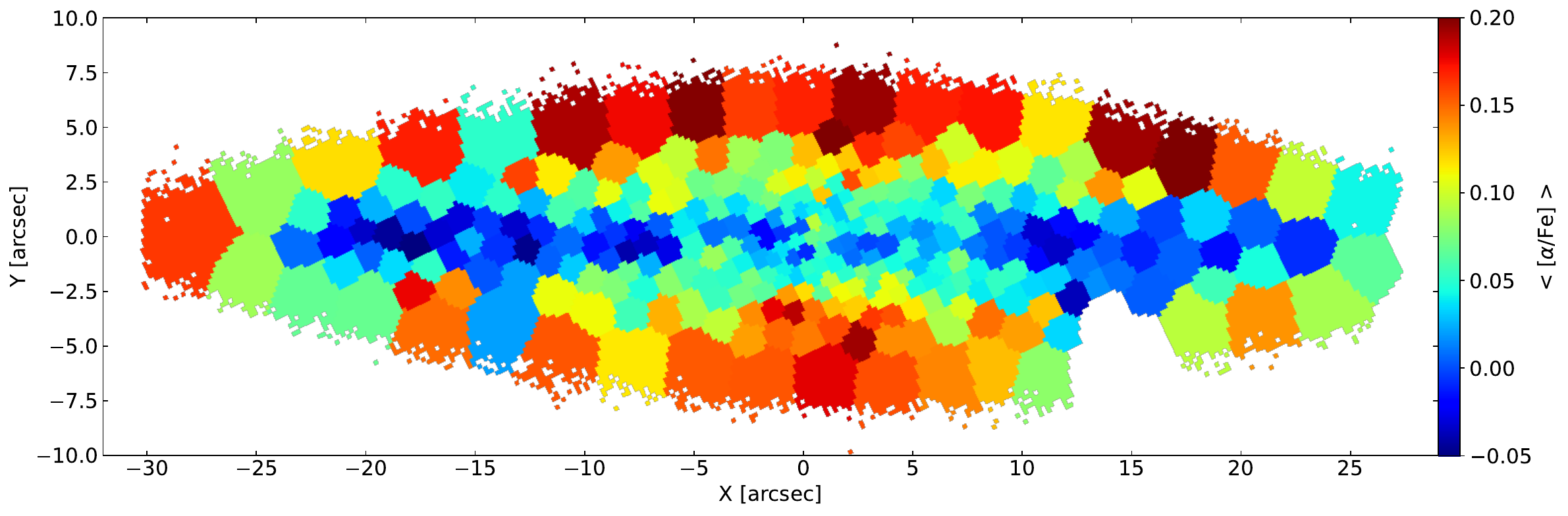}
    \caption{Spatial distribution of mass-weighted mean age (top), [M/H] (middle), and [$\alpha$/Fe] (bottom) for each spatial bin in ESO~544-27.}
    \label{Fig: mean_maps}
\end{figure*}

\section{ESO 544-27 and its IFU Spectroscopy}
\label{sect: obs}

ESO~544-27 (PGC 8480) is an Sb type galaxy \citep[][]{Sorce14} at a distance of $\sim45.9$~Mpc \citep{Tully16}. The stellar mass of the galaxy is $1.1 \times 10^{10} \rm~M_{\odot}$ \citep {Comerón2018}, making it less massive than the MW ($5.4 \times 10^{10} \rm~M_{\odot}$; \citealt {McMillan2017}). It is an emission-line galaxy with a significant population of hot low-mass evolved stars in the halo \citep{RAUTIO}. The star-formation rate (SFR) of the galaxy calculated using H$\alpha$ luminosity is around $0.13~\rm M_{\odot}/yr$, while its specific-SFR (sSFR) is $10^{-10.7} \rm /yr$, making it a green-valley galaxy \citep{RAUTIO}. ESO~544-27 also has significant dust extinction in the central region, and a prominent dust lane in the galaxy's mid-plane can be seen in Figure~\ref{Fig: ESO 544-27}.  

\citet {Comerón2018} found that thin and thick discs can fit the vertical structure of ESO~544-27 using 3.6 $\mu m$ images from the S4G survey. They found the thin disc has a scale height of $\sim0.7''$ (0.16~kpc), while the thick disk has a scale height of $\sim3''$ (0.67~kpc). They estimated that the mass of the thin and thick discs are $7.6 \times 10^{9} M_{\odot}$ \& $3.6 \times 10^{9} M_{\odot}$ respectively. There is no clear signature of a bulge in the galaxy's central region.

MUSE has a spatial sampling of $0.2''$ with a $1'\times 1'$ field-of-view and wavelength range from 4750--9351~\AA \citep{2010Bacon}. It has a spectral resolution 2.5~\AA. ESO~544-27 had four 2624~s dithered exposures and the combined reduced data-cube was made available through the ESO Phase 3 data-release\footnote{\url{https://doi.eso.org/10.18727/archive/8}}.  The details of the observations \& their reduction are mentioned in \citet {Comerón2019}. The MUSE field-of-view of ESO 544-27 has two contaminating galaxies  (marked in red in Figure~\ref{Fig: ESO 544-27}) as well as some foreground stars/ globular clusters. We used the ESO~544-27 MUSE white light images to create a mask to remove the contaminants. 


\begin{figure*}
\centering
\includegraphics[width=\textwidth]{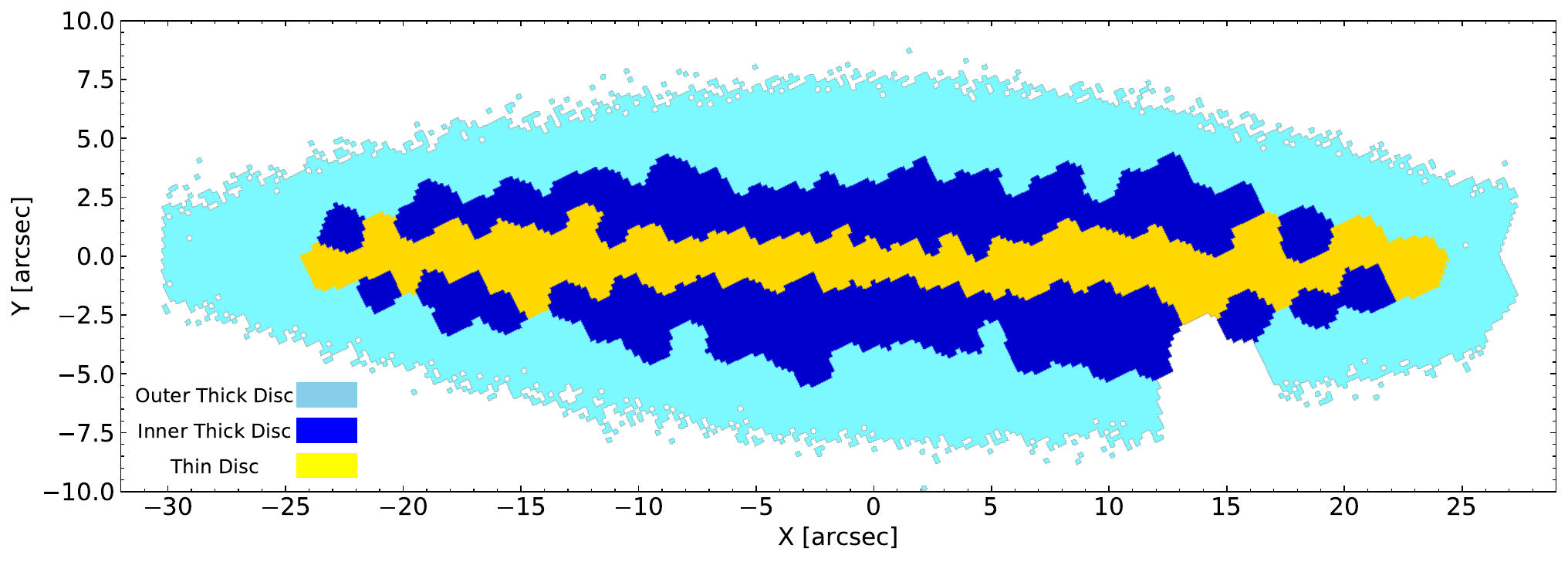}
\caption{The spatial bins of ESO~544-27 classified into different regions.}
\label{Fig: regions}
\end{figure*}

\begin{figure}
    \centering
    
    \includegraphics[width=\columnwidth]{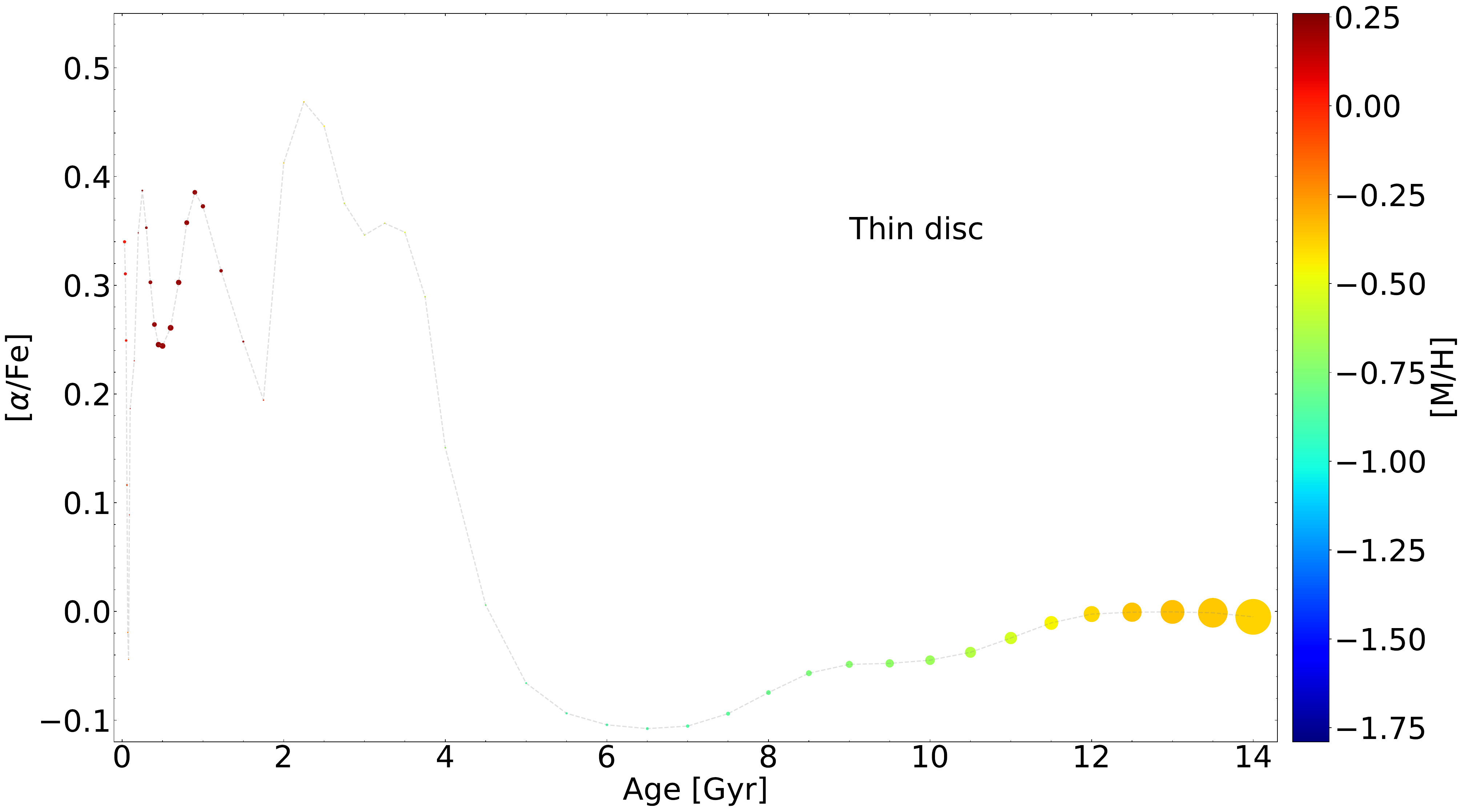}
    
    
    \includegraphics[width=\columnwidth]{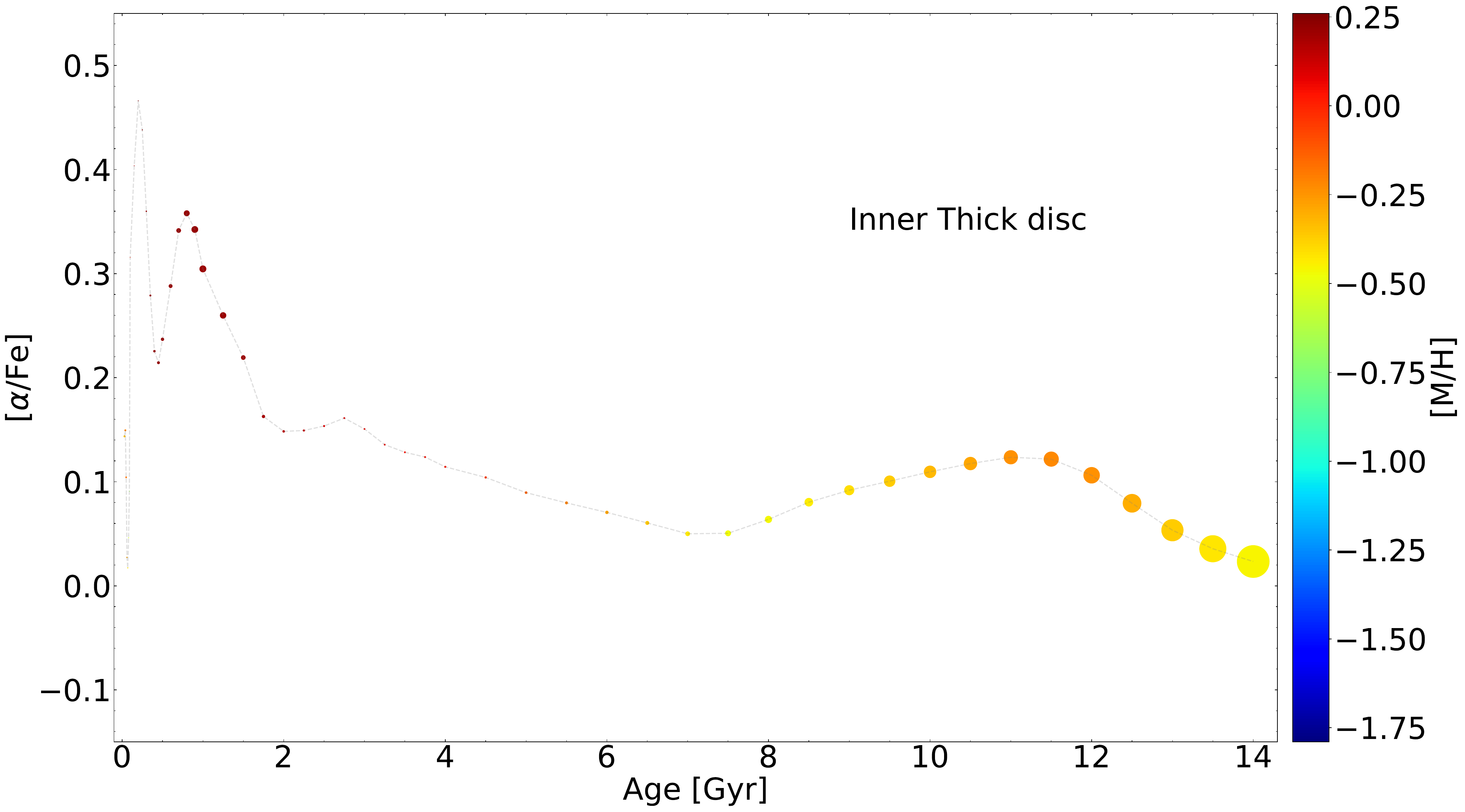}
    
    
    \includegraphics[width=\columnwidth]{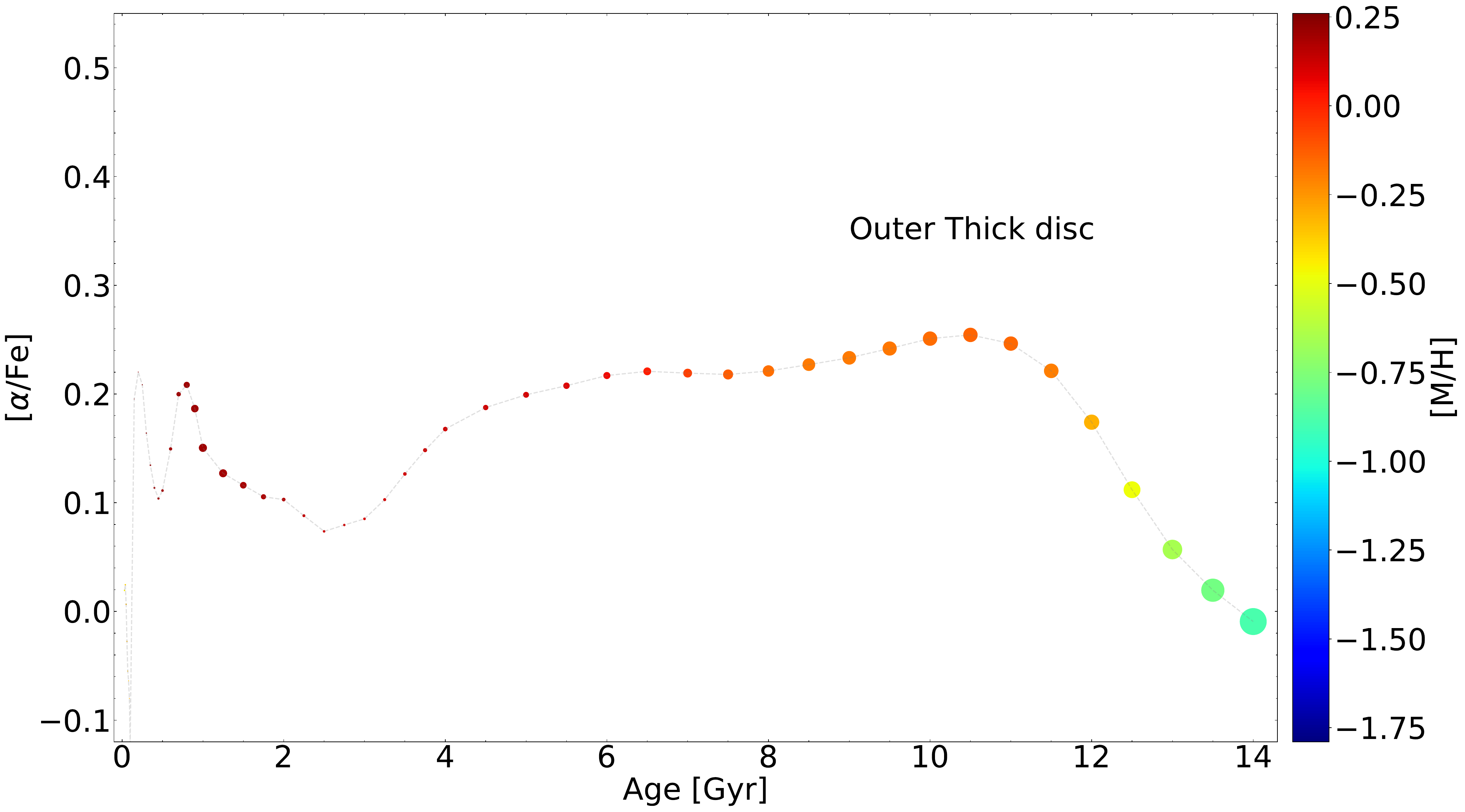}

    \caption{The mass-weighted average of [$\alpha$/Fe] in different age bins is plotted against the stellar age of each bin, separately for the stellar populations in the thin [top], inner thick [middle], and outer thick [bottom] disc of ESO~544-27. Each data point is coloured by the mass-weighted mean [M/H] for each age bin and its diameter is proportional to the average mass fraction of all the spatial bins in that age bin.}
    \label{Fig: alpha}
    
\end{figure}

\section{Stellar population Analysis}
\label{sect: analysis}

From the masked MUSE cube for ESO 544-27, we carried out voronoi-binning \citep{Cappellari03} of the spectral cube to obtain spatially binned-spectra of adequate signal-to-noise (S/N) to allow for stellar population model fitting. This was accomplished using the GIST (Galaxy IFU Spectroscopy Tool) Pipeline \citep{BITTNER}. We adopted a minimum S/N of 60 per spatial bin in the spectral range of 4800--5500~\AA. For the S/N ratio per spaxel, we kept 3 as a minimum, as used in \citet{BITTNER}. This resulted in a total of 304 spatial bins for the galaxy.

For the spectra of each spatial bin in ESO 544-27, we extract stellar population properties by full spectrum fitting using pPXF (Penalized Pixel-Fitting method) described in \citet{Cappellari2004}, and upgraded in \citet{Cappellari2017}. Stellar kinematics and stellar population model fitting are carried out simultaneously from the observed spectra. This method employs Gauss-Hermite series expansions of the line-of-sight velocity distribution to fit the observed spectra to templates from a stellar spectral model library. We utilised the Semi-empirical (sMILES) single stellar population (SSP) models spectra. These are available at the MILES \citep{Vazdekis15} sampling, resolution (FWHM), and wavelength range levels of 0.9~\AA, 2.5~\AA, and 3540.5--7409.6~\AA, respectively. This library contains 5 values for [$\alpha$/Fe] abundances ranging from $-$0.2 to +0.6 in 0.2 steps, 10 values of metallicity [M/H] = [$-1.97$, +0.26] dex and 53 values of total age from 0.03 and 14.0 Gyr. The SSP models number 2650 in total. Compared to the MILES SSP models (having [$\alpha$/Fe]= 0.0 \& 0.4) which were used in similar earlier works \citep{2019PinnaA,2019PinnaB,2021MMarie,2021Scott,2023Satler}, the sMILES most notably allows higher resolution in [$\alpha$/Fe] abundances. We have used unimodal Salpeter IMF \citep{Salpeter55} with a logarithmic slope of 1.3. We found that assuming a Kroupa IMF \citep{Kroupa01} makes little difference to the determined mass fractions.

As ESO 544-27 is an emission line galaxy with some residual star-formation \citep{RAUTIO}, it is necessary to mask emission lines that affect the stellar population model fitting with ppxf. We masked H-$\beta$, H-$\alpha$, [SII]$\lambda$6717,6731, [OIII]$\lambda$4959,5007, [OI]$\lambda$6300, and [NII]$\lambda$6548,6583. The stellar population of the spectra was then fitted (in the spectral range of 4800--7371~\AA, where MUSE observations and sMILES SSP models overlap) by fixing the multiplicative polynomial to 8th order as was deemed appropriate to account for the mid-plane dust lane in ESO 544-27, similar to FCC~170 \citep{2019PinnaA}. The use of additive polynomials in stellar populations was avoided to prevent changes in the absorption lines, which could have affected the values of [$\alpha$/Fe] and metallicity \citep{2016Guerou}.

The spectra of each bin is first dust corrected. The stellar extinction ($A_{v}$) is obtained from fitting each spectra without regularisation using pPXF. Each spectra is then dust corrected with the $A_{v}$ value thus obtained assuming the \citet{Calzetti94} dust attenuation law. The spatial distribution of $A_{v}$ values is shown in Figure~\ref{Fig: Dust_correction}.

To obtain the best fit for the dust-corrected spectra of each bin, we regularised the stellar population weights assigned to the single stellar population (SSP) models \citep{Cappellari2017}. We choose a fixed regularization parameter value of 0.07 (similar to 0.05 for \citealt{2019PinnaA}). This was chosen by finding a balance between smoothing the solutions to remove noise and not losing star-formation history information. The procedure followed was similar to that carried out by \citet{2019PinnaA} and \citet{2021Scott}. The best-fit model spectra for a spatial bin is further discussed in Appendix~\ref{App: bestfit}. We thus obtained the mass weights for each SSP model using ppxf for all 304 spatial bins of ESO~544-27. We removed four bins due to poor fits or contamination from foreground stars. We also applied bootstrapping to the pPXF full spectral fitting of each bin to obtain uncertainty on the mass weights following \citet{Cappellari23}, described in Appendix~\ref{App: boot}.

We note that the stellar populations properties derived by pPXF for a given spectra may still have some biases from the true stellar population properties. \citet{Woo24} found that for spectra with S/N $\sim$ 20, stellar age may be overestimated by pPXF (while utilising the E-MILES spectral library, \citealt{Vazdekis16}) by $\sim0.5$~Gyr for ages lower than $\sim$3~Gyr though older ages have negligible bias. They also found that [M/H] may be overestimated by $\sim$0.07~dex at solar metallicities, with higher bias of upto $\sim$0.25~dex for lower [M/H] values. E-MILES did not have [$\alpha$/Fe] as a free parameter. Given our significantly higher S/N of 60 per spatially binned spectra and use of the sMILES library, we expect lower biases in the determined stellar population properties. But until a similar analysis to \citet{Woo24} is carried out for the sMILES library at higher S/N, we can expect a maximum bias in our determined ages of $\sim0.5$~Gyr for younger ages, and of $\sim$0.25~dex for low [M/H] values.

\section{Results}
\label{Sect: Results}


\subsection{Mass-weighted mean stellar population properties}
\label{sect: mean_maps}

The mass weights were used to obtain the mean age, [M/H], and [$\alpha$/Fe] for each spectral bin in ESO~544-27. Their spatial distribution is plotted in Figure~\ref{Fig: mean_maps}. ESO~544-27 has an almost uniformly old stellar population with an age $\sim10$~Gyr for most spatial bins. Different structures in the galaxy can be seen in the age spatial distribution (Figure~\ref{Fig: mean_maps} [top]), although the age difference is small. The metallicity spatial distribution (Figure~\ref{Fig: mean_maps} [middle]) also shows small differences in the average between the different structures in the galaxy. It shows that the spatial bins close to the galactic mid-plane are relatively metal-rich (mean [M/H] > -0.4), though some bins have lower metallicities. However, the spatial bins away from the mid-plane appear to have even lower metallicities (mean [M/H] < -0.45).

However, a clear transition in the galaxy structures can be observed in the [$\alpha$/Fe] spatial distribution (Figure~\ref{Fig: mean_maps} [bottom]). The galaxy mid-plane region spatial bins have [$\alpha$/Fe]$\sim-$0.05--0.05, while those at larger scale-heights have mean [$\alpha$/Fe]$\sim$ 0.05--0.125, increasing to mean [$\alpha$/Fe]$\sim$ 0.1--0.2 furthest from the mid-plane. No clear bulge structure is seen at the galaxy's central regions in either of the panels of Figure~\ref{Fig: mean_maps}. The low [$\alpha$/Fe] mid-plane is consistent with the low velocity-dispersion thin disc identified in ESO~544-27 by \citet{Comerón2019} while the higher scale-height high [$\alpha$/Fe] regions likely correspond to their identified higher velocity-dispersion thick disc.

The mean age, [M/H], and [$\alpha$/Fe] spatial distributions do not provide a complete understanding of the galaxy's stellar population properties. However, they provide sufficient insight into the galactic structure to explore further diagnostics for the thin and thick discs of ESO~544-27. Figure~\ref{Fig: mean_maps} [bottom] provides a clear guideline to distinguish between the different disc structures in ESO~544-27 from the mean [$\alpha$/Fe] values of its spatial bins.

The thin disc-dominated region, having mean [$\alpha$/Fe]$\leq$ 0.025, lies in the range of -24$''$ < X < 24$''$ and -1.5$''$ < Y < 1.5$''$. The spatial bins in this region are marked in yellow in Figure~\ref{Fig: regions}. While the remaining spectral bins may be classified into the thick disc, we account for the possibility that a transition zone may be present between the thin and thick discs, which may not be consistent with a clear boundary solely based on mean [$\alpha$/Fe]. We have divided spectral bins above and below the mid-plane with mean [$\alpha$/Fe] $\geq0.1$ value into the outer thick disc. The remaining spectral bins have been marked as inner thick disc (at lower scale heights), as shown in Figure~\ref{Fig: regions}. The thin disc region contains 90 spatial bins, while the inner and outer thick disc regions contain 154 and 56 spatial bins respectively.


    

    

    




\begin{figure*}
    \centering 
\begin{subfigure}{\columnwidth}
  \includegraphics[width=\textwidth]{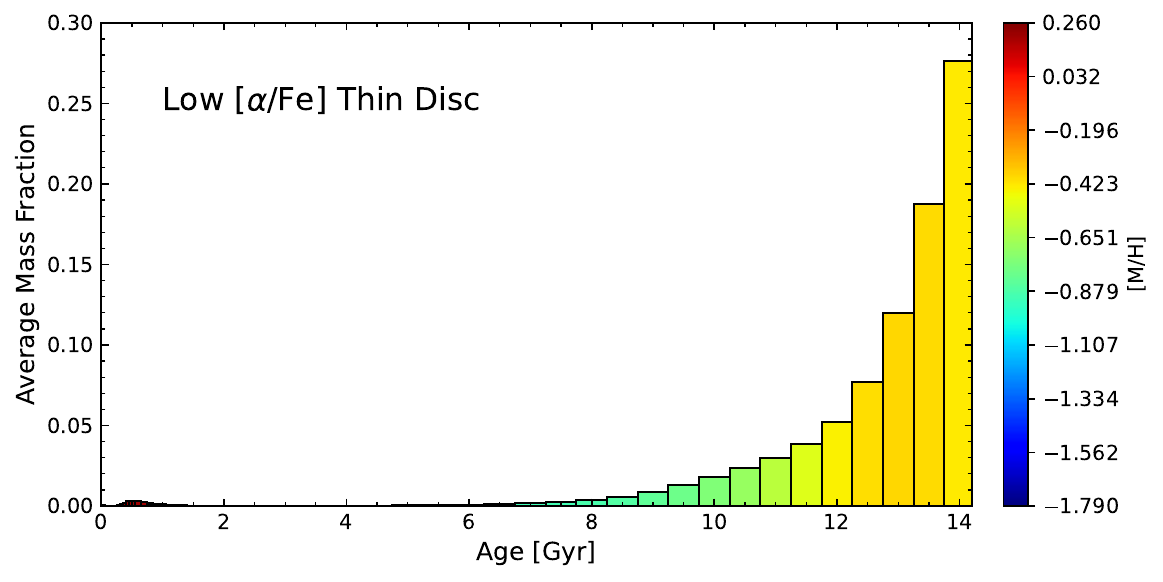}
\end{subfigure}\hfil 
\begin{subfigure}{\columnwidth}
  \includegraphics[width=\textwidth]{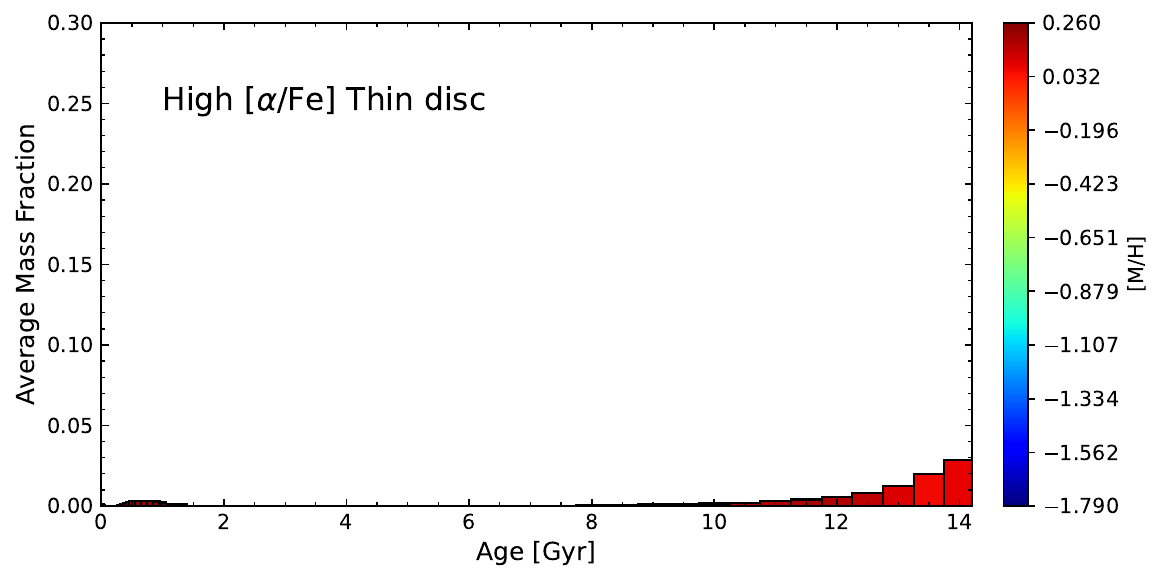}
\end{subfigure}\hfil 

\begin{subfigure}{\columnwidth}
  \includegraphics[width=\textwidth]{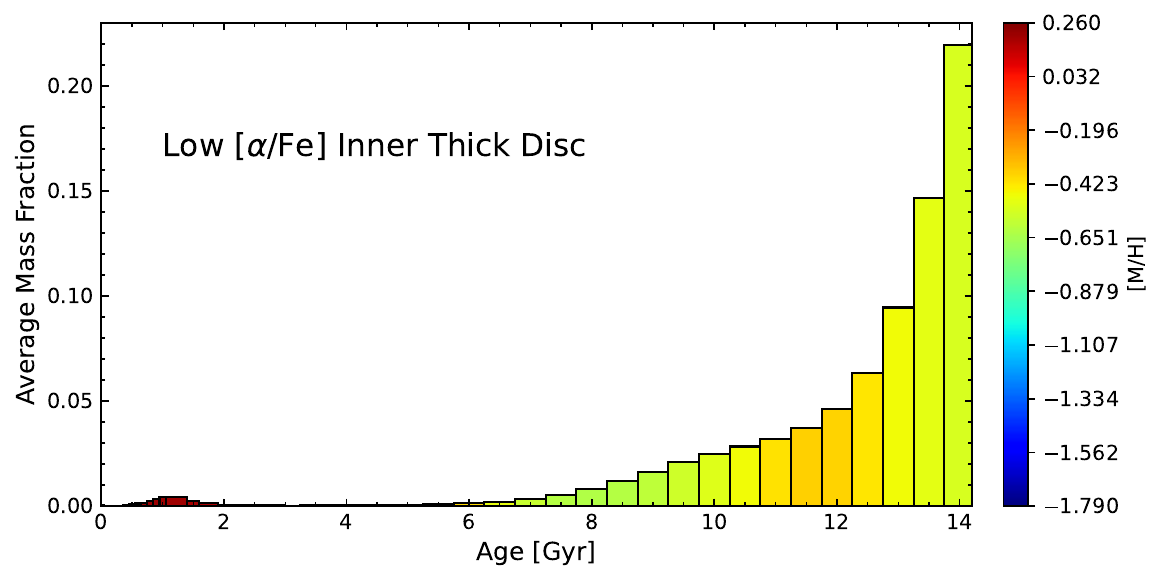}
\end{subfigure}\hfil 
\begin{subfigure}{\columnwidth}
  \includegraphics[width=\textwidth]{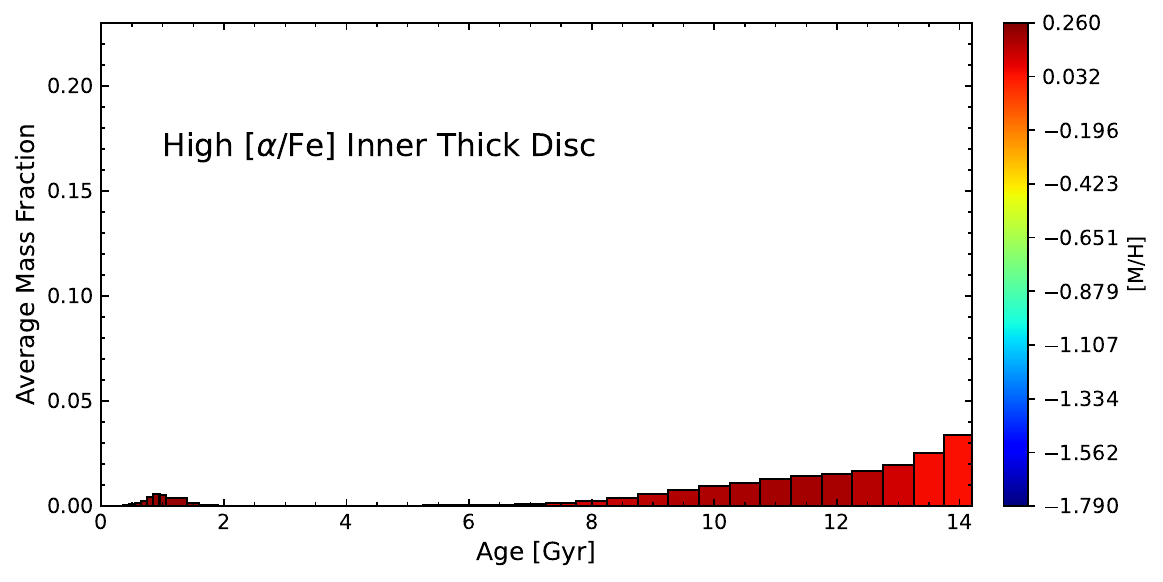}
\end{subfigure}\hfil 

\begin{subfigure}{\columnwidth}
  \includegraphics[width=\textwidth]{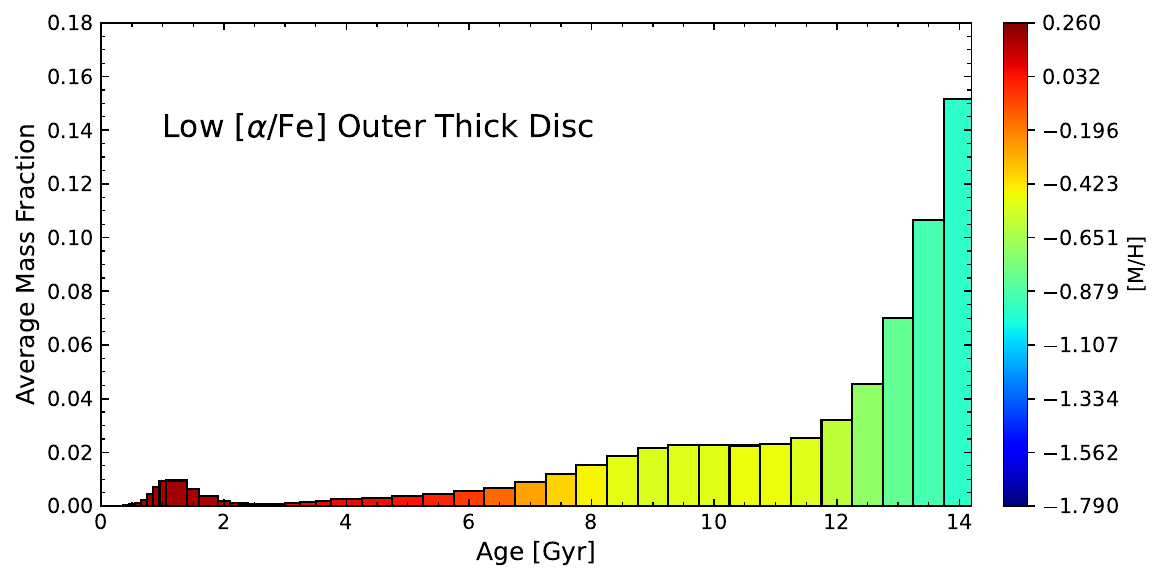}
\end{subfigure}\hfil 
\begin{subfigure}{\columnwidth}
  \includegraphics[width=\textwidth]{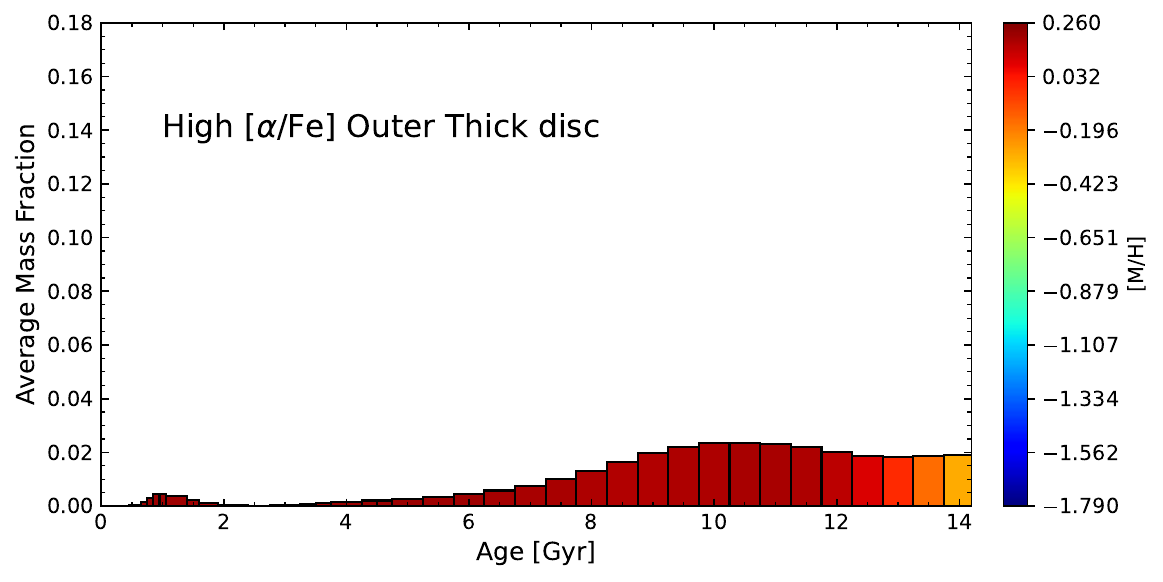}
\end{subfigure}\hfil 

\caption{Average mass fraction is plotted as a function of age in each age bin, coloured by its mass-weighted mean [M/H], with the thin [top], inner thick [middle], and outer thick [bottom] disc stellar populations are divided into low $[\alpha$/Fe] [left] (-0.2 -- 0.2) \& high $[\alpha$/Fe] [right] (0.4 -- 0.6) samples.}
\label{fig:images}
\label{Fig: bar_mass}
\end{figure*}


\begin{figure}
    \centering
    
    \includegraphics[width=0.8\linewidth]{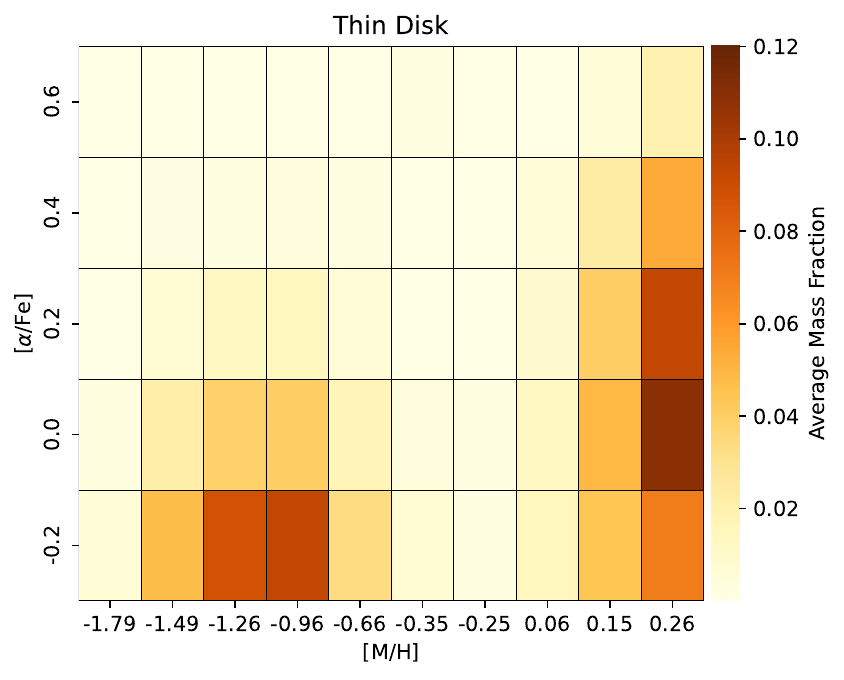}
    
    
    \includegraphics[width=0.8\linewidth]{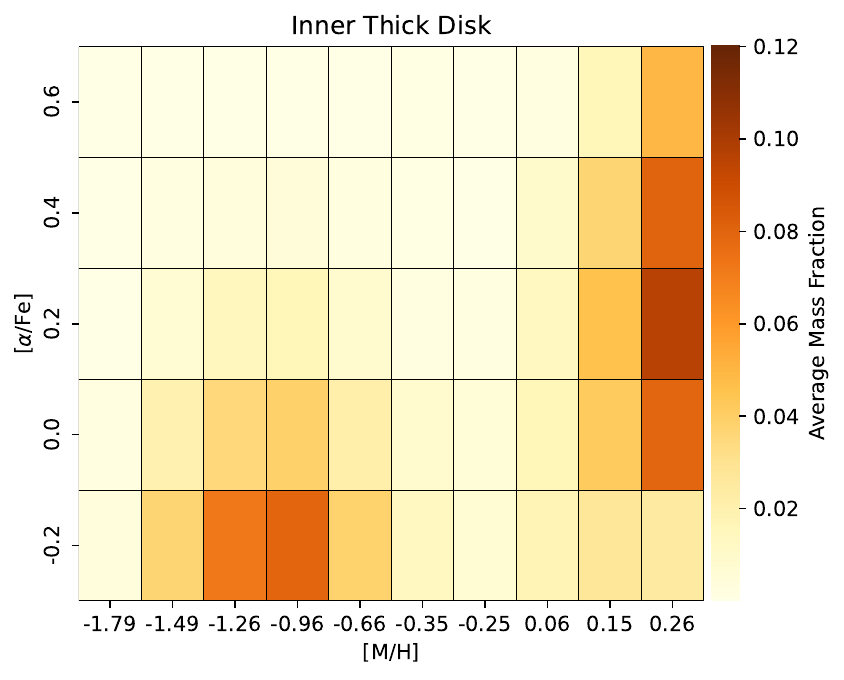}
    
    
    \includegraphics[width=0.8\linewidth]{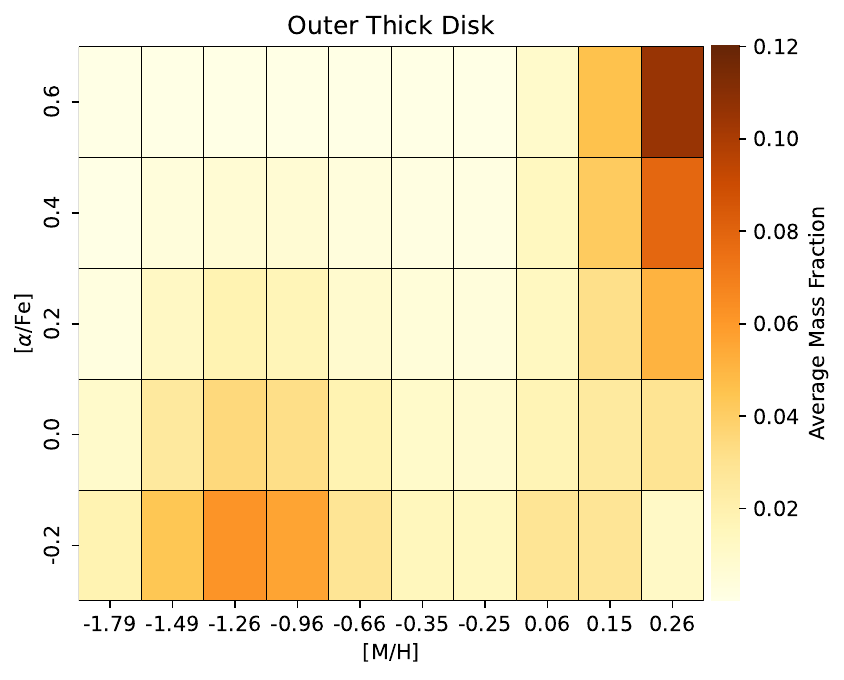}

    \caption{The [$\alpha$/Fe] vs [M/H] plane for stellar populations of all ages in the thin [top], inner thick [middle], and outer thick [bottom] disc regions of ESO~544-27. The plane is binned as per the [$\alpha$/Fe] and [M/H] grid described in Section~\ref{sect: analysis} with each grid region coloured by the determined average mass fraction.}
    \label{Fig: population}
    
\end{figure}

    

\begin{figure*}
    \includegraphics[width=\textwidth]{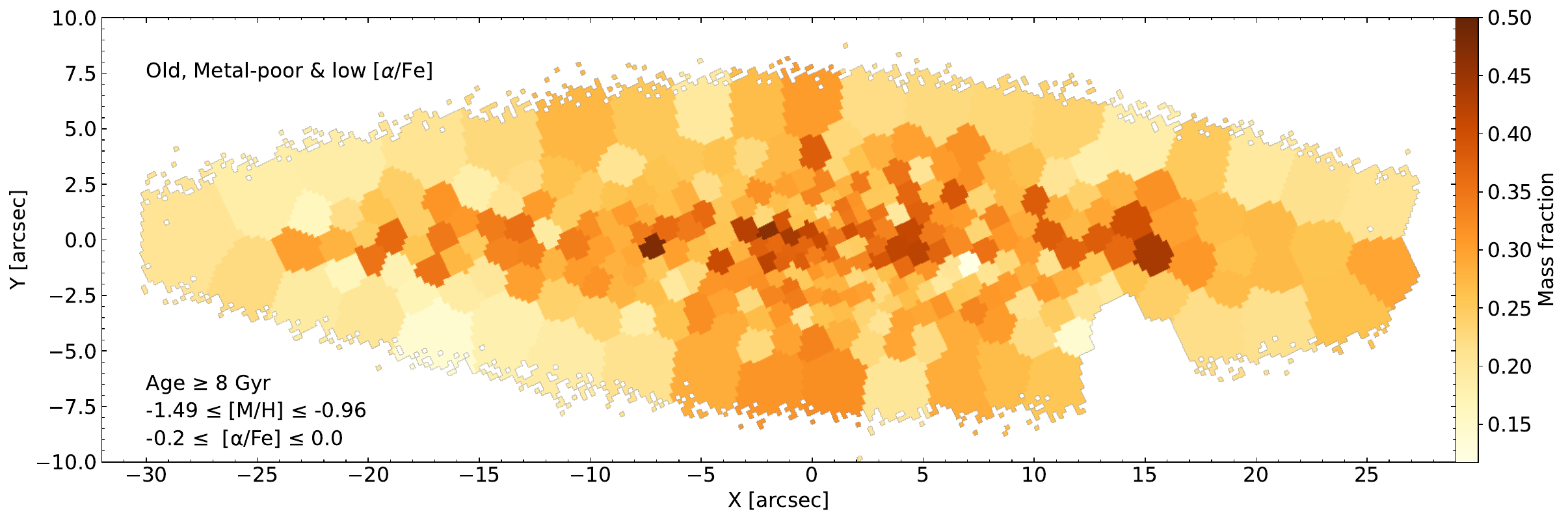}
    
    
    \includegraphics[width=\textwidth]{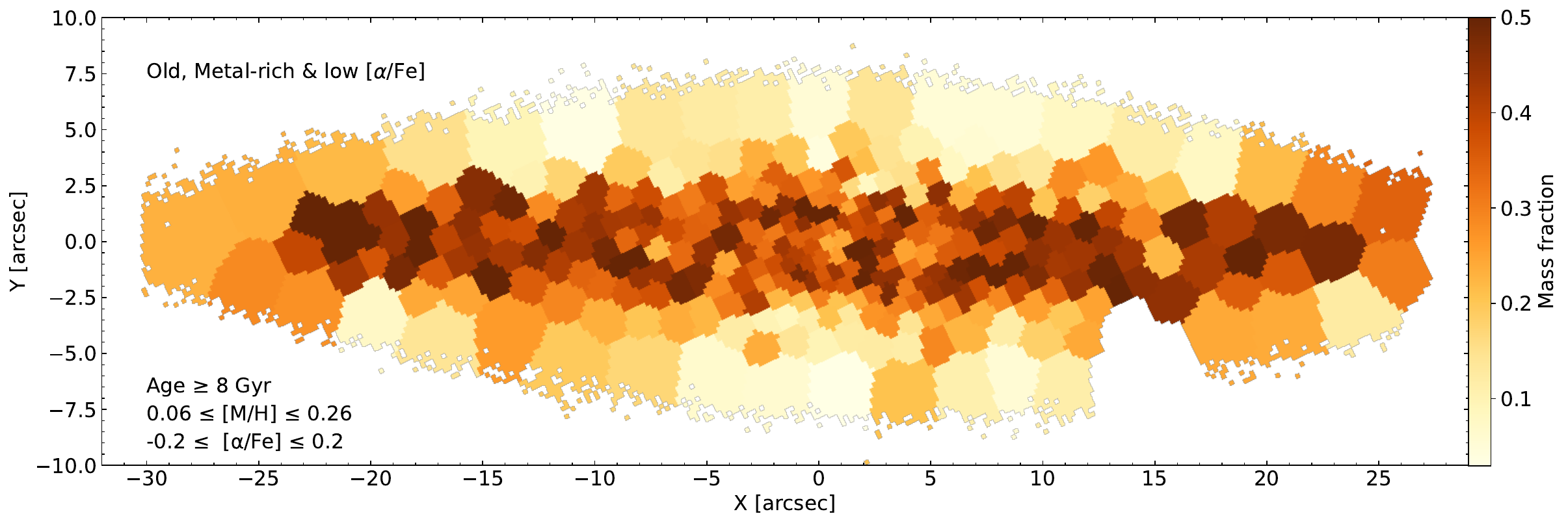}
    
    
    \includegraphics[width=\textwidth]{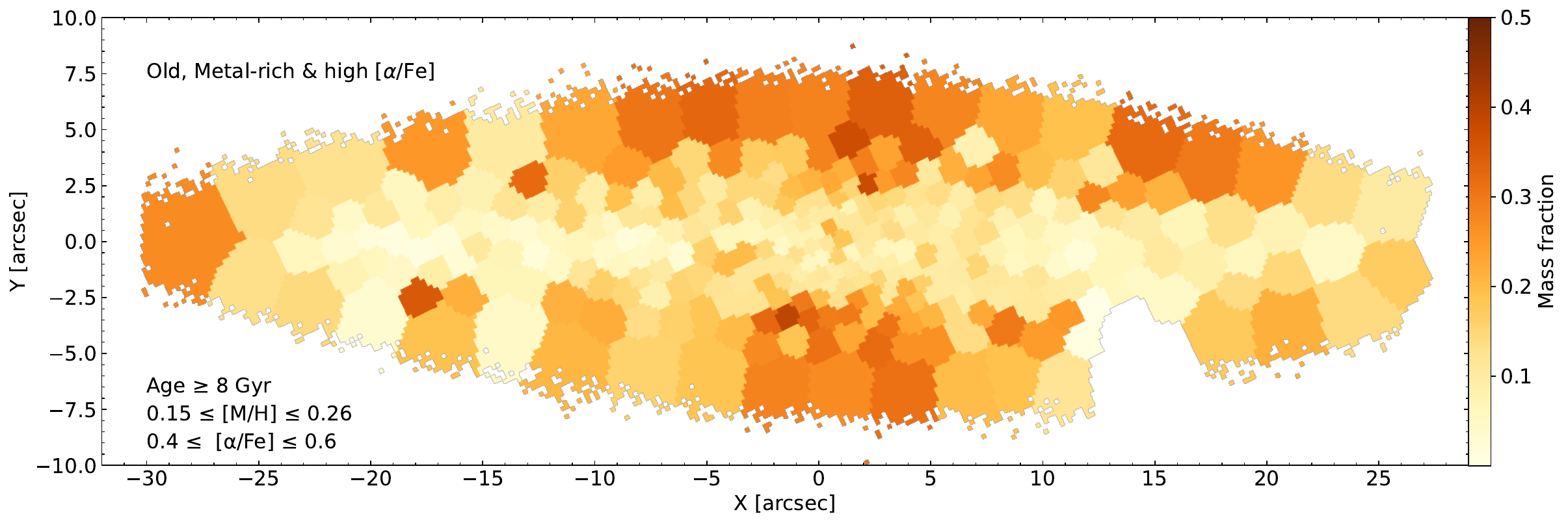}
    \caption{The spatial distribution of the old ($\geq$8~Gyr) stellar populations in ESO~544-27 that is metal-poor and low [$\alpha$/Fe] (top), metal-rich and low [$\alpha$/Fe] (middle), and metal-rich and high [$\alpha$/Fe] (bottom). }
    \label{Fig: old_populations}
\end{figure*}

\subsection{Stellar population properties in the galaxy disc regions binned in age} 
\label{sect:disc_age}

Figure~\ref{Fig: alpha} plots stellar age against the mass-weighted average of [$\alpha$/Fe] abundances for each age bin (the 53 age-bins resolved by the sMILES library; see Section~\ref{sect: analysis}; note the age-bins are wider for older ages than for younger ones), separately for the stellar populations in the thin, inner thick, and outer thick disc of ESO~544-27. The figure is colour-coded by the mass-weighted mean [M/H] for each age bin. The range of [M/H] is -1.79 -- 0.26. The diameter of each data point in the figure is proportional to the average mass fraction of all the spatial bins in that age bin. The average mass fraction summed over the age bins in each region is unity.

For the thin disc of ESO~544-27 (Figure~\ref{Fig: alpha} [top]), the stellar population aged 12--14~Gyr old has low [$\alpha$/Fe] with [M/H] $\sim-0.36$. The stellar population in this age range has the highest average mass fraction relative to all younger populations in the thin disc. Between the ages of 8--12~Gyr, the thin disc stellar population also has low [$\alpha$/Fe] but is relatively metal-poor with [M/H] $\sim -0.66$. The average mass fraction is lower for stars born in this age range and shows a clear decreasing trend for younger ages. There is no significant average mass fraction of thin disc stars born in the age range of 2–8 Gyr. For the age-range of 0.3--1.25~Gyr, a metal-rich ([M/H]>0) higher [$\alpha$/Fe] population is present in the thin disc.  

For the inner thick disc of ESO~544-27 (Figure~\ref{Fig: alpha} [middle]), the stellar population aged 12--14~Gyr old shows a mildly increasing trend in [$\alpha$/Fe] and [M/H] with decreasing age, peaking in both [$\alpha$/Fe] and [M/H] for the 12~Gyr old stellar population. Like the thin disc, the highest average mass fraction of the inner thick disc was formed in this age range. Between the ages of 8--12~Gyr, the stellar population is higher [$\alpha$/Fe] than the thin disc population of the same age, similarly for metallicity with [M/H] $\sim -0.37$. The average mass fraction is lower in this age range for the inner thick disc as was the case for the thin disc, again showing a clear trend of decreasing average mass fraction for younger ages. There is more average mass fraction in the inner thick disc than the thin disc for the age range of 2--8~Gyr, but still the average mass fraction is very small but relatively metal-rich ([M/H]>0).  For the age-range of 0.5--1.75~Gyr, a metal-rich ([M/H]>0) high [$\alpha$/Fe] population similar to that in the thin disc is also present in the inner thick disc.  

For the outer thick disc of ESO~544-27 (Figure~\ref{Fig: alpha} [bottom]), the stellar population aged 11--14~Gyr old shows a strongly increasing trend in [$\alpha$/Fe] and [M/H] with decreasing age, peaking in both [$\alpha$/Fe] and [M/H] for the 11~Gyr old stellar population. The 14~Gyr old population, in particular, is much more metal-poor ([M/H]$\sim$-0.9) than the similarly aged population in the thin and inner thick disc regions. Again, most of the average mass fraction in the inner thick disc formed 11-14 Gyr ago. The stellar population then has a nearly constant  [$\alpha$/Fe]$\sim 0.24$ till the age decreases to $\sim$8~Gyr, whereupon [$\alpha$/Fe] start increasing \& [M/H] remains nearly constant but with reducing average mass fraction. The average mass fraction formed ~2-3 Gyr ago is negligible. For the age range of 0.7--2~Gyr, a metal-rich ([M/H]>0) high [$\alpha$/Fe] population similar to that in the thin and inner thick disc is also present in the outer thick disc, but with higher average mass fraction.  

Each panel of Figure~\ref{Fig: bar_mass} depicts average mass fraction as a function of age in each age bin, coloured by its mass-weighted mean [M/H], but now the thin, inner thick, and outer thick disc stellar populations are divided into low $[\alpha$/Fe] \& high $[\alpha$/Fe] samples with $[\alpha$/Fe]= -0.2 -- 0.2 \& 0.4 -- 0.6 respectively. For the thin, inner thick \& outer thick disc, $\sim$88\%, $\sim$79\% \& $\sim$68\% of the mass respectively is low [$\alpha$/Fe], while the rest is high [$\alpha$/Fe]. 

Thus the high-[$\alpha$/Fe] average mass fraction increases as we move away from the galaxy's mid-plane. The high [$\alpha$/Fe] mass is almost exclusively older than 12~Gyr in the thin disc, while it's only almost exclusively older than 8~Gyr in the inner thick disc. The high [$\alpha$/Fe] population persists at almost all ages in the outer thick disc though the average mass fraction increases to 8 Gyr and then starts decreasing. The high [$\alpha$/Fe] population is predominantly metal-rich ([M/H]>0), with the exception of the oldest stellar population in the outer thick disc, which exhibits [M/H] around $\sim-0.3$.

The low [$\alpha$/Fe] stellar population is as described for Figure~\ref{Fig: alpha}, the highest average mass fraction in each regions formed 12--14~Gyr ago. There is a clear decrease in metallicity for this age range as we move away from the mid-plane with the thin disc being most metal-rich with [M/H]$\sim -0.41$, the inner thick disc having [M/H]$\sim -0.46$, and outer thick disc having [M/H]$\sim -0.8$. The average mass fraction in the age range of 2--8~Gyr old increases as we move away from the mid-plane, though it remains very low everywhere.

The younger stellar population of 0.5--2~Gyr old age is persistently metal-rich at all regions, though it forms a relatively higher average mass fraction in the outer thick disc. This population seems split between low and high [$\alpha$/Fe].


\subsection{\texorpdfstring{[$\alpha$/Fe] vs [M/H] in the galaxy disc regions}{[alpha/Fe] vs [M/H] in the galaxy disc regions}}
\label{sect:disc_alpha}

Figure ~\ref{Fig: population} depicts average mass fraction over all spatial bins of the thin, inner thick, and outer thick disc regions in the [$\alpha$/Fe] vs [M/H] plane for stellar populations of all ages. For each region, sum of the average mass fraction depicted in this plane will be unity. In all three regions, we find a stellar population having [$\alpha$/Fe] $\leq0$ and [M/H]= -1.49 -- -0.96. All three regions also have substantial metal-rich stellar populations ([M/H]>0.15), but each region differs in the [$\alpha$/Fe] where most of this metal-rich mass is present. 

For the thin disc, the metal-rich population is mostly low [$\alpha$/Fe] with [$\alpha$/Fe]$\leq0.2$ while for the outer thick disc, the metal-rich population is mostly high [$\alpha$/Fe] with [$\alpha$/Fe]$\geq0.4$. The inner thick disc has the metal-rich population spanning a wider [$\alpha$/Fe] range with [$\alpha$/Fe]= 0 -- 0.6. There is thus a transition from low to high [$\alpha$/Fe] for the metal-rich population as we move away from the mid-plane of the galaxy from the thin disc to the outer thick disc.


\subsection{The spatial distribution of the old stellar population}
\label{sect: spatial_old}

ESO~544-27 has a mostly old stellar population that has a metal-poor low [$\alpha$/Fe] component as well as a metal-rich component with varying [$\alpha$/Fe]. Figure~\ref{Fig: old_populations} shows the spatial distribution of the old ($\geq$8~Gyr old) stellar populations in ESO 544-27. 

In Figure~\ref{Fig: old_populations} [top], we see the spatial distribution of the metal-poor low [$\alpha$/Fe] stellar population. The low [$\alpha$/Fe] metal-poor stellar population is present throughout the galaxy. However, since it is the only low [$\alpha$/Fe] stellar population present in the outer thick disc (see Section~\ref{sect:disc_alpha}), it makes the overall outer thick disc appear relatively metal poor and contributes to the overall transition of the low [$\alpha$/Fe] stellar population from metal-rich to metal-poor as we move away from the mid-plane (Figure~\ref{Fig: bar_mass}).
We note here that such a low [$\alpha$/Fe] metal poor stellar population is not expected for older stellar populations which is expected to instead have a high [$\alpha$/Fe] metal poor component \citep[e.g.][]{Vincenzo18}. 

We ascertain that the detection of this population arises
from the limited ability of pPXF for stellar population model fitting in the presence of dust attenuation. This population is present with the highest mass fraction in the midplane region of ESO 544-27 (see Figure~\ref{Fig: old_populations} [Top]) also featuring a prominent dust lane (Figure~\ref{Fig: ESO 544-27}). This population was also observed by \citet{2021Scott} at low-scale heights in UGC~10738 (see their Figure 5). In Appendix~\ref{App: dust}, we correct for dust attenuation from the Balmer decrement in the spectra of a single spatial bin before employing full spectral fitting with pPXF. While this does reduce the mass fraction attributed to this old low [$\alpha$/Fe] metal-poor stellar population, the mass fraction change is within error and the mean age, metallicity and [$\alpha$/Fe] values of the other stellar populations remain nearly unaffected. We further discuss the effect of dust attenuation in Appendix~\ref{App: dust}. In Appendix~\ref{App: bestfit}, we find that restricting our fit for the spectra of a single bin to only high [$\alpha$/Fe] stellar populations, still provides for an acceptable fit. We thus exclude this old low [$\alpha$/Fe] metal-poor stellar population from the analysis.

In Figure~\ref{Fig: old_populations} [middle], we see the spatial distribution of the metal-rich low [$\alpha$/Fe] stellar population ([$\alpha$/Fe]$\leq$0.2). This population is present almost entirely in the mid-plane of the galaxy. In Figure~\ref{Fig: old_populations} [bottom], we see the spatial distribution of the metal-rich high [$\alpha$/Fe] stellar population ([$\alpha$/Fe]$\geq$0.4). As was expected from Figure~\ref{Fig: population}, this population is noticeably absent in the galaxy mid-plane and starts appearing at larger scale-heights, mostly present in the outer thick disc region.

\subsection{The spatial distribution of the younger stellar population}
\label{sect: spatial_young}

\begin{figure*}
    
\includegraphics[width=\textwidth]{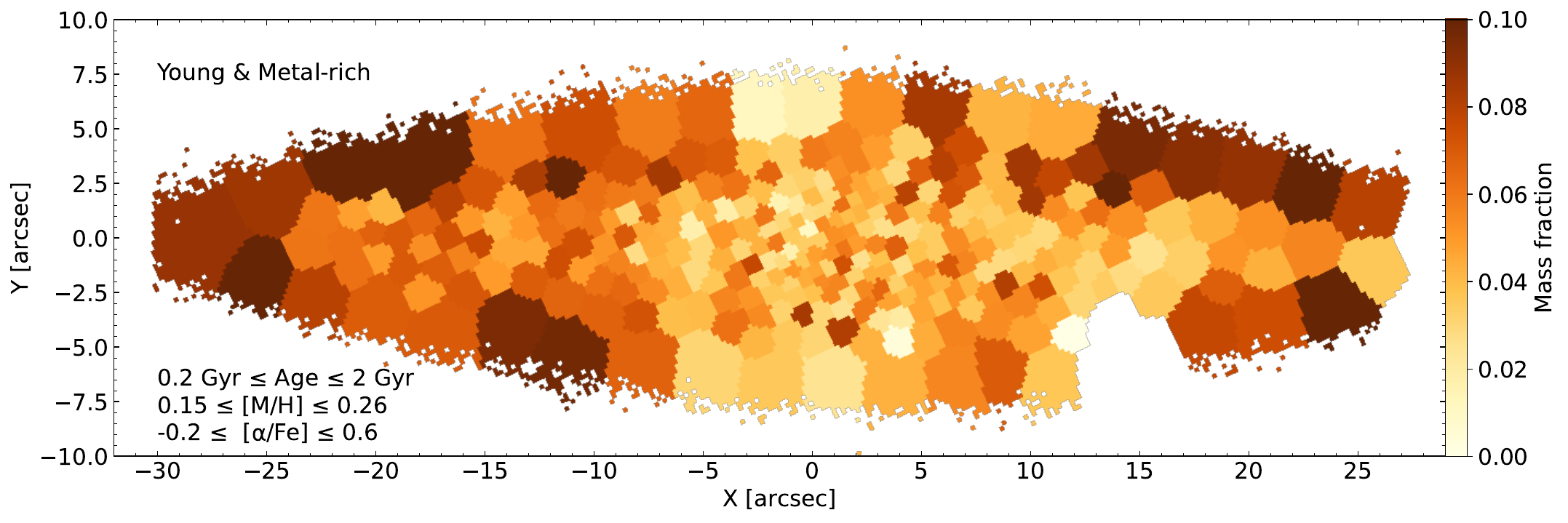}
\caption{The spatial distribution of the metal-rich younger stellar populations in ESO~544-27. The mass fraction range is appropriately chosen for this stellar population.}
\label{Fig: young_populations}
\end{figure*}

ESO~544-27 has a relatively young metal-rich stellar population present, as was evident in Figures~\ref{Fig: alpha}~\&~\ref{Fig: bar_mass}.  Its spatial distribution is plotted in Figure~\ref{Fig: young_populations}. The age of this population ranges from 0.2 Gyr to 2 Gyr while the range of metallicity is 0.15 dex $\le$ [M/H] $\le$ 0.26 dex. This population spans the entire range of [$\alpha$/Fe]. White it forms only $\sim$1.2\% of the total stellar population of the galaxy by mass, it is present in the outer thick disc spatial bins with its highest mass fraction of $\sim0.1$ and with lower mass fractions in the inner thick and thin discs.


\begin{figure*}
    \centering
    \includegraphics[width=0.32\textwidth]{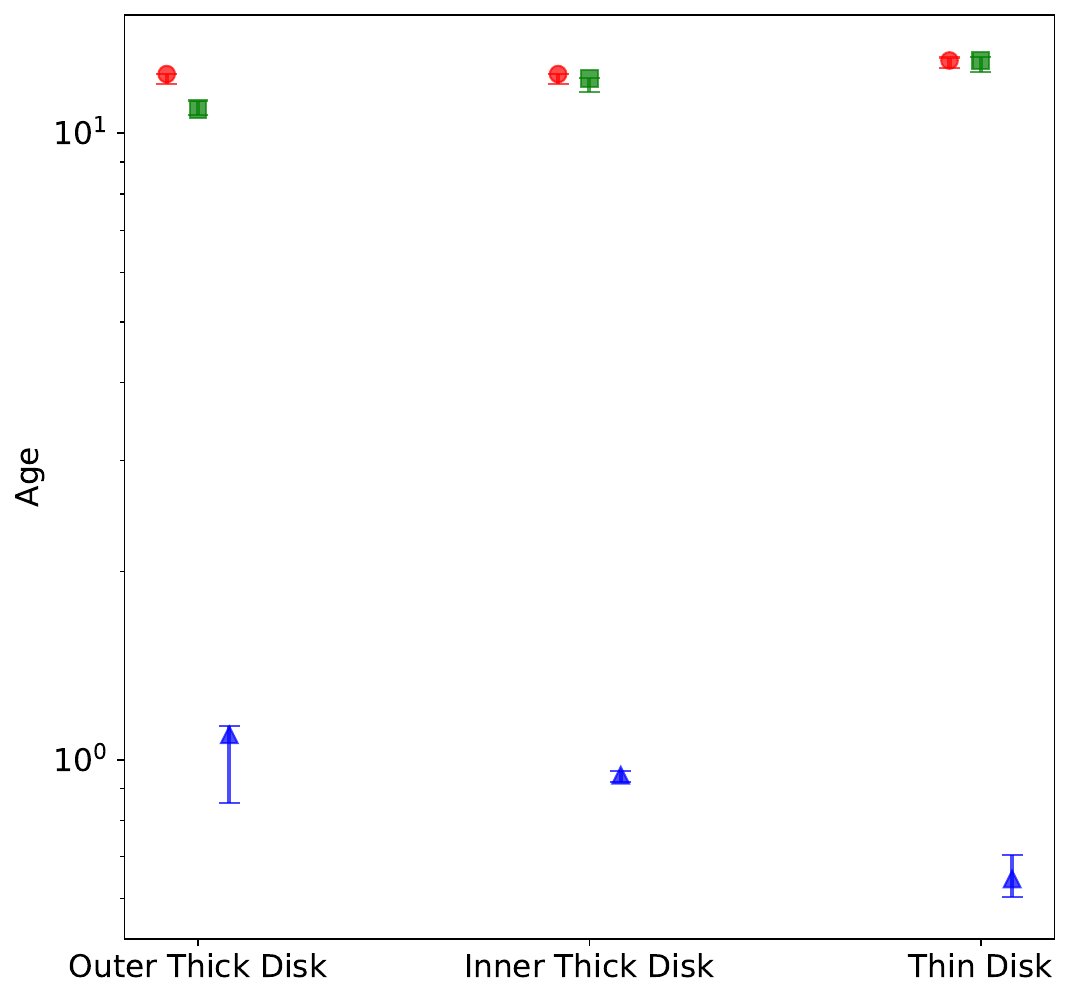}
    \includegraphics[width=0.32\textwidth]{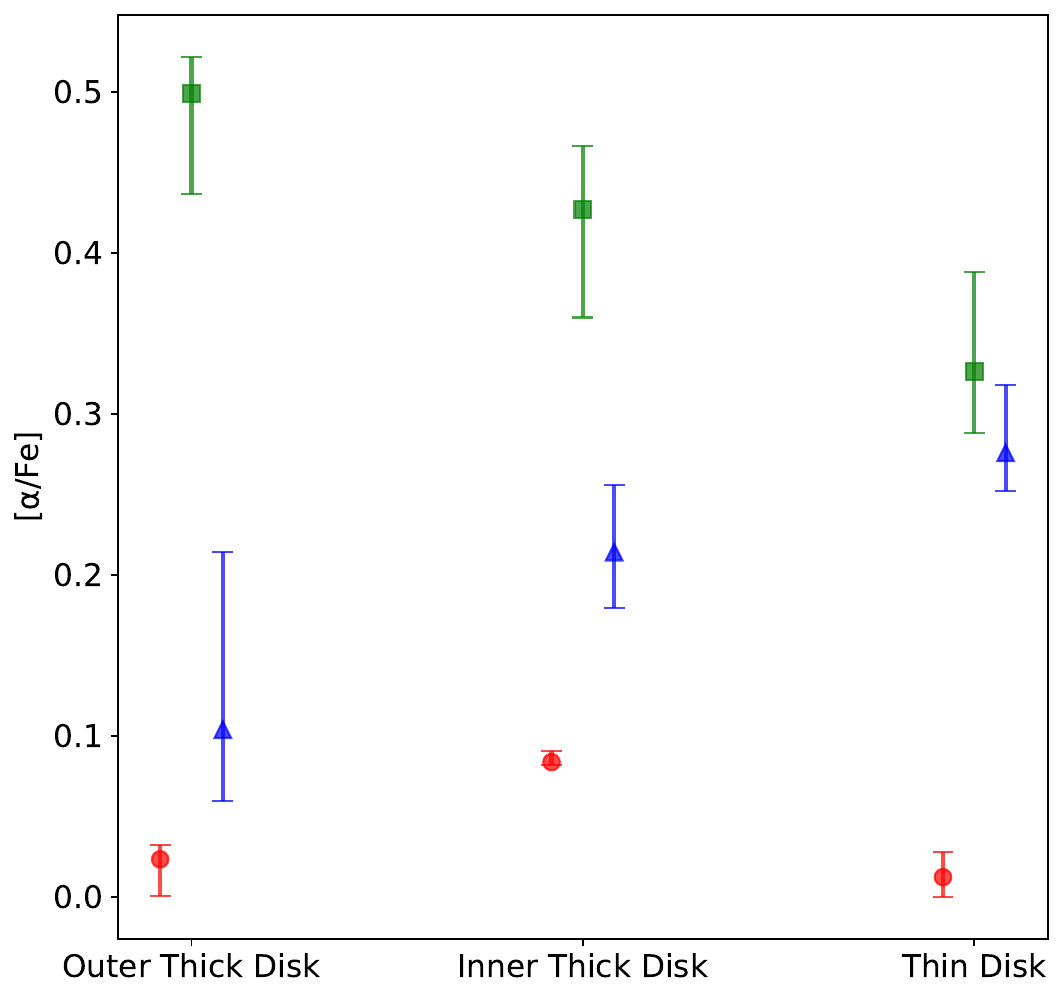}
    \includegraphics[width=0.325\textwidth]{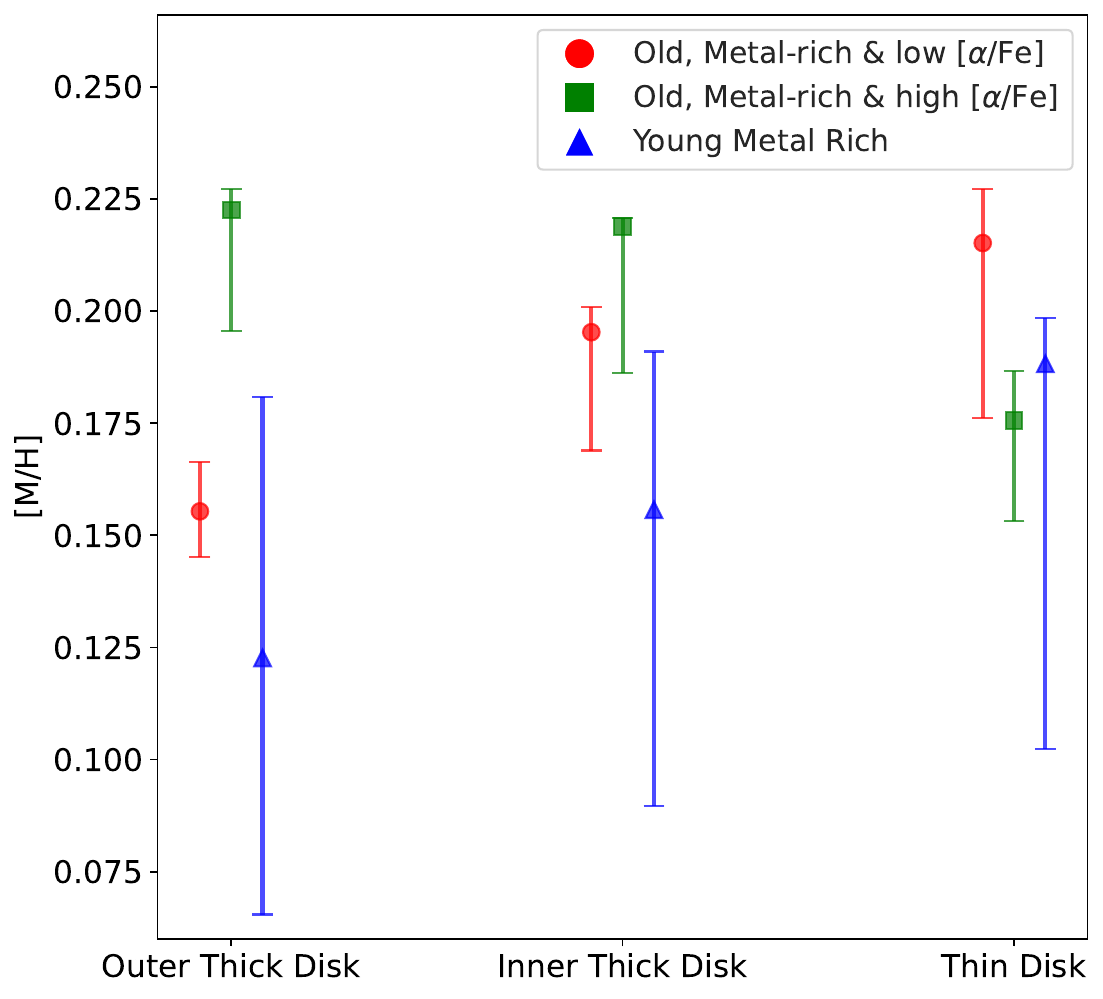}
    
    \caption{Mean properties (age [left], [$\alpha$/Fe] [center], [M/H] [right]) of the stellar populations (old, metal-rich and low [$\alpha$/Fe] -- red; old, metal-rich and high [$\alpha$/Fe] -- green; younger metal-rich -- blue) in the three regions of ESO 544-27.}
    \label{Fig:Error Bar}
    
\end{figure*}

\subsection{Mean properties of the different stellar populations} 
\label{sect:disc_mean}

In Figure~\ref{Fig:Error Bar}, we show for the three different regions, the mean age [left], [$\alpha$/Fe] [center] and [M/H] [right] for the old, metal-rich and low [$\alpha$/Fe] stellar population, the old, metal-rich and high [$\alpha$/Fe] stellar population, and the younger metal-rich stellar population. The errors associated with the mean properties for each population were determined through bootstrapping the full spectral fitting method with ppxf. The bootstrapping method is discussed in Appendix~\ref{App: boot}. 

The mean age of the younger metal-rich population is lower in the thin disc. Any bias in the determined age from pPXF, up to $\sim$0.5~Gyr, is expected to affect all younger ages equally while no effect is expected for the older ages \citep{Woo24}. A rising trend is seen for the [$\alpha$/Fe] of this population as we move from outer thick disc to thin disc, while a decreasing trend is seen for the [$\alpha$/Fe] of the old, metal-rich and high [$\alpha$/Fe] stellar population. All populations have mean super-solar metallicities. We note however that pPXF may overestimate the [M/H] value at low metallicities by up to $\sim0.25$~dex \citep{Woo24}. See also Section\ref{sect: analysis} for discussion on the expected biases.


\section{Discussion}
\label{sect: discussion}


\subsection{The origin of the thin and thick disc stellar populations}
\label{sect: origin}

From the diagnostic results presented in Section~\ref{Sect: Results}, we consequently try to understand the origin of the stellar populations is ESO~544-27. High [$\alpha$/Fe] stellar population is expected to be formed in rapid star-formation events where enrichment from core-collapse supernovae dominates while low [$\alpha$/Fe] stellar populations correspond to more extended star-formation events where there has been sufficient time for Type Ia supernovae enrichment to occur \citep[e.g.][]{Kobayashi20}. Efficient star formation can increase metallicity, even up to supersolar values (see bulge model in \citealt{Kobayashi20}).

In ESO~544-27, the bulk of the stellar population is old with the high [$\alpha$/Fe] stellar population mainly present in the outer thick disc (Figure~\ref{Fig: old_populations} [bottom]). The high [$\alpha$/Fe] stellar population likely formed in-situ reaching the most metal-rich stellar populations very quickly close to the mid-plane and forming less metal-rich stellar populations away from the mid-plane for the oldest ages ($\geq13$~Gyr; see Figure~\ref{Fig: bar_mass}). Younger ($<$12~Gyr old) high [$\alpha$/Fe] stellar populations were all metal-rich. 

The low [$\alpha$/Fe] metal rich stellar population is present mainly in the thin disc of the galaxy (Figure~\ref{Fig: old_populations} [middle]). This population likely formed over an extended period though the ISM had become metal-rich fairly quickly (Figure~\ref{Fig:Error Bar}). The galaxy thin disc displays a low velocity dispersion \citep{Comerón2019}, thus appearing unperturbed from major mergers. Thus this stellar population likely formed in the galaxy thin disc and is observed mostly in-situ. 

The youngest stellar population in ESO~544-27 is metal-rich, present mainly in the outer thick disc (Figure~\ref{Fig: young_populations}).  This population is younger with a higher [$\alpha$/Fe] in the thin disc. If this population had formed in the thin disc and had later been ejected outwards, we would have expected similar ages everywhere. Additionally as ESO~544-27 had minimal star-formation in all spatial regions between 2--8~Gyr ago (Figure~\ref{Fig: bar_mass}), it was almost quenched and thus this stellar population likely comes from an ex-situ origin. Gas infall from external galaxies, unaccompanied by stars, would have resulted in the gas settling and forming stars only in the thin disc. Even if these would be later ejected out, the same age would have been observed for young stars in all three regions.

Through N-body simulations of mergers, \citet {2019Karademir} found that in minor-mergers, satellite stars are deposited at higher scale-heights \citep[see also][]{Read08}. Depending on the orbit, an infalling gas-rich dwarf galaxy experiences a burst of star-formation upon entering the outer thick disc of the galaxy \citep{Zhu23}. Most stars of the dwarf would be stripped and deposited at the outer thick disc, and while the older stars in the dwarf would be more difficult to distinguish from the stars present in the outer thick disc prior to the merger, the newly formed stars would stand out in the star-formation history (Figure~\ref{Fig: bar_mass}). 

As the satellite moves further into the galaxy at lower scale heights, less material is deposited but it can be traced that the newly-formed stars deposited in the inner thick and thin disc are all progressively younger on average than those deposited in the outer thick disc (Figure~\ref{Fig:Error Bar}). The younger stellar population is seemingly consistent with forming in a wet minor merger event. 

However, such metal-rich ([M/H]$>$0) stellar population is not typically observed in dwarf galaxies (see the mass-metallicity relation in \citealt{Hidalgo17}). Some metal-rich dwarf galaxies have been observed, having relatively low gas fractions and nearing the end of their star formation activity \citep{Peeples08}. It is possible that such a dwarf galaxy's infall is responsible for the formation of the younger metal-rich stellar population. Such dwarfs are also expected to have a significant amount of older stars but those would have been dispersed (primarily in the outer thick disc) in this scenario and thus rendered indistinguishable from the older stars of ESO 544-27. Alternatively, tidal-dwarf galaxies (TDGs) can also reach similarly high metallicities and are offset from the mass-metallicity relation of most dwarf galaxies \citep[e.g.][]{Recchi15}. Thus it is also possible that the cannibalized dwarf was a TDG. Other alternative scenarios may be possible to explain the origin of the younger population in ESO~544-27 but such scenarios would also need to account for the difference in mean ages measured for the distinct disc regions for this population.

We thus find that for the thick disc of ESO~544-27, multiple pathways of thick disc formation have contributed. {The old high [$\alpha$/Fe] stellar population either formed at its position in the thick disc, or formed in the thin disc and was expelled to larger scale heights at later times.} The younger metal-rich population likely comes from a wet minor merger event.

For the thin disc of ESO~544-27, the old high [$\alpha$/Fe] stellar population is present. This may have formed in-situ in the thin disc of the galaxy as some early high [$\alpha$/Fe] stellar population is expected to form at early times from galactic chemical evolution models \citep[e.g.][]{Kobayashi20}. However, this population forms only a small fraction of the mass of the thin disc. The old low [$\alpha$/Fe] metal-rich stellar population also likely formed from in-situ gas at early times and inhabits low scale heights till present times. Finally the younger population also inhabits the thin disc, likely coming from a wet minor merger event but forming a very small fraction of the mass.


\subsection{ESO 544-27 in comparison to other disc galaxies}
\label{sect: context}

The MW thin disc has a significantly younger stellar population in comparison to its thick disc \citep[e.g.][]{GilmoreReed1983,2013Haywood}. So do UGC 10378 \citep{2021Scott} and NGC 3501 \citep{2023Satler}. Even in most external galaxies, the thin disc has generally been considered to be younger than the thick disc \citep[e.g:][]{2016Kasparova,2008Yoachim,2016Guerou,2015Comeron}. This is very different from ESO~544-27, which has an overall older population, including in the thin disc. The thin disc of ESO~544-27 is similar to that of the S0 galaxy FCC~170 \citep{2019PinnaA} as well as those of NGC~5422 \citep{2016Kasparova} and ESO 243-49 \citep{Comeron2016}, wherein the thin disc is dominated by old metal-rich, low [$\alpha$/Fe] stellar populations. The thin disc stellar population being more metal-poor at ages $\sim$8--12~Gyr is also similar to that observed for NGC 3501 \citep{2023Satler}. This was interpreted as indicating an early infall of metal-poor gas. However, as we see in this work, the mean-metallicity is affected by the pPXF-identified old likely unphysical metal-poor low [$\alpha$/Fe] stellar populations, while the physical old low [$\alpha$/Fe] stellar populations is actually metal-rich.

The MW has a old high [$\alpha$/Fe], relatively metal-poor thick disc \citep[e.g.][]{2015Hayden}.  So do UGC 10378 \citep{2021Scott}. This is in contrast to the thick disc of ESO~544-27 which while consistent with having high [$\alpha$/Fe] stellar population, has a rather metal-rich one. The thick disc of ESO~544-27 likely has the high [$\alpha$/Fe] component formed higher star-formation efficiency that the MW so as to reach higher metallicities. This is similar to the old M~31 thick disc stellar population that is also similarly more metal-rich \citep{Kobayashi23}. 
 
{The younger stellar population present mostly in the thick disc and little in the thin disc of ESO~544-27 most likely comes from a wet minor merger event $\sim1$~Gyr ago though it forms a very small fraction of the mass.} From the distribution of stellar population ages (Figure~\ref{Fig: bar_mass}), ESO~544-27 was nearly, if not completely, quenched between $\sim$2--8~Gyr ago with negligible star-formation. Whatever little star-formation takes place in the galaxy at present times, likely originates from the gas brought in by this wet minor merger event. \citet{RAUTIO} calculated the specific star formation rate of ESO 544-27 using the extinction corrected H$\alpha$-luminosity, which comes out to be the least in their analysed sample of five edge-on galaxies from \citet{Comerón2019}, placing the galaxy green valley of star formation. We may thus picture ESO~544-27 as a reignited S0 galaxy.
\section{Conclusion}
\label{sect: conclusion}

Using deep MUSE integral-field spectroscopy, we have carried out an analysis of the stellar populations of the edge-on galaxy ESO~544-27. We studied the spatial distribution of the mass-weighted mean stellar age, [M/H] and [$\alpha$/Fe]. We found that the galaxy is dominated by old stellar populations (mean age $\sim$10~Gyr). Further diagnostics (Sections~\ref{sect: analysis}~\&~~\ref{Sect: Results}) revealed the following:
\begin{itemize}
    \item The thin disc of ESO~544-27 is dominated by an old low [$\alpha$/Fe] metal-rich stellar population, that likely formed in-situ.
    \item The outer thick disc has a high [$\alpha$/Fe] metal-rich component (also present with a smaller mass fraction in the thin disc) that was likely formed in-situ but with higher star-formation efficiency than the MW thick disc.
    \item The inner thick disc of this galaxy has stellar population properties in-between its thin and outer thick disc.
    \item We thus find [$\alpha$/Fe] dichotomy in ESO 544-27 with its thin and thick discs dominated by low and high [$\alpha$/Fe] stellar populations respectively.
    \item The galaxy was nearly (if not completely) quenched between $\sim$2--8~Gyr ago but star-formation was reignited recently, first in the outer and inner thick disc ($\sim$1 Gyr ago) and then in the thin disc ($\sim$600 Myr ago). This is consistent with being induced by a wet merger event $\sim$1~Gyr ago with a dwarf galaxy but such a metal-rich stellar population is uncommon in dwarfs.
\end{itemize}

We thereby demonstrate the significant potential of stellar population model-fitting of IFU observations of edge-on galaxies to provide insight on the formation pathways of thin and thick discs, revealing the diverse ways in which they may form. In particular ESO~544-27 has a unique origin compared to the disc galaxies where such a study has been conducted (Section~\ref{sect: discussion}). Similar analysis is planned for the larger sample of edge-on galaxies with MUSE observations by \citet{Comerón2019}.


\section*{Acknowledgements}
We thank the anonymous referees for their comments. SB is funded by the INSPIRE Faculty Award (DST/INSPIRE/04/2020/002224), Department of Science and Technology (DST), Government of India. The award also supported DS during his recurring stays at IUCAA, Pune, India. We acknowledge the use of Pegasus, the
high-performance computing facility at IUCAA. This research made use of Astropy-- a community-developed core Python package for Astronomy \citep{Rob13}, SciPy \citep{scipy}, NumPy \citep{numpy} and Matplotlib \citep{matplotlib}. This research also made use of NASA’s Astrophysics Data System (ADS\footnote{\url{https://ui.adsabs.harvard.edu}}).

\section*{Data Availability}
The reduced MUSE data-cube for ESO 544-27 is publicly available through the ESO Phase 3 data-release at \url{https://doi.eso.org/10.18727/archive/8}.


\bibliographystyle{mnras}
\bibliography{example} 

\appendix
\section{}
\subsection{Best-fit stellar population for a spatial bin}
\label{App: bestfit}

\begin{figure*}
\includegraphics[width=\textwidth]{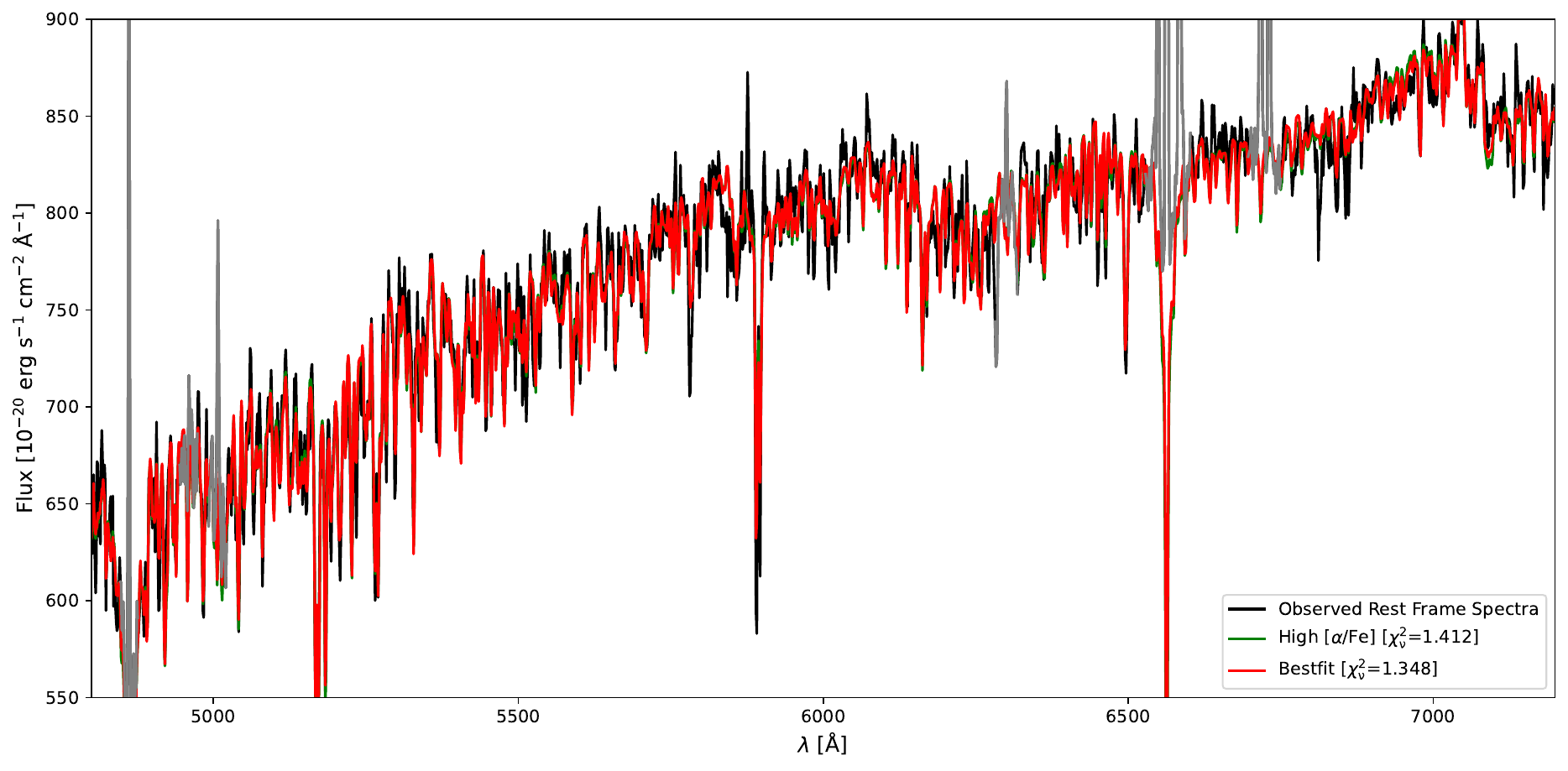}
\caption{Examples of stellar population model fits for the spectra of a single spatial bin. The observed spectrum (shifted to the rest-frame) is shown in black (with masked emission lines in grey). The best-fit spectrum considering all sMILES stellar population models, including the old [M/H]-rich and low-[$\alpha$/Fe] stellar population, is marked in red. The best-fit spectrum considering only the high-[$\alpha$/Fe] stellar population models is marked in green. The $\rm\chi^{2} _{\nu}$ values for the fits are noted.}
\label{Fig: appendix}
\end{figure*}

Figure~\ref{Fig: appendix} shows the rest-frame observed spectra for a single spatial bin and examples of its stellar population model fits. Our fitting procedure (Section~\ref{sect: analysis}) produces the best-fit spectra. Its $\rm\chi^{2} _{\nu}$ is the lowest (note the value is off from one due to regularisation done in ppxf). The best-fit for the spectra in most spatial bins of this galaxy leads to the presence of a low [$\alpha$/Fe], metal-poor component (Section~\ref{sect: spatial_old}). As such a component may not be present physically, we check the fitting with only high [$\alpha$/Fe] models. This too produces a reasonable fit though the best-fit model, with a lower $\rm\chi^{2} _{\nu}$ value, is still preferred. Finally, as an old metal-poor population may produce spectra that are similar to those of a younger metal-rich stellar population, we check the fitting with only younger ($\leq2$~Gyr old) models. This produces a very bad fit with $\rm\chi^{2} _{\nu} = 39.49$ (not shown in Figure~\ref{Fig: appendix}), indicating that ESO~544-27 has very little contribution from younger stellar populations, as was also observed by \citet{RAUTIO}. 

\subsection{Estimating errors in full spectral fitting with Bootstrapping}
\label{App: boot}

Bootstrapping is a methodology that enables estimating the sampling distribution for nearly any statistic by employing random sampling techniques \citep{Efron93}. It directly explains variance, bias and other probabilistic phenomena. We are implementing bootstrapping to obtain errors in the estimated mass weights of different stellar populations fitted by pPXF following the methodology from \citet{Cappellari23}. Repeated ppxf fits, combined with bootstrapping of the residuals, enable the calculation of both averages and uncertainties in the distribution of the weights. For each galaxy spatial bin spectrum, we generate a set of 30 realizations, with the understanding that increasing the number of realizations does not affect the results (for some spatial bins, we checked larger number of realizations and saw negligible changes in the mass weight errors). Subsequently, we apply the same measurement technique to each simulated set. The measurements on each simulated set are carried out using the same initial parameters as those used to measure the original spectrum. We calculated errors of mass weights for each bin using this approach.

\subsection{Estimating the effect of dust attenuation correction}
\label{App: dust}

\begin{figure*}
    \centering
    \includegraphics[width=0.7\columnwidth]{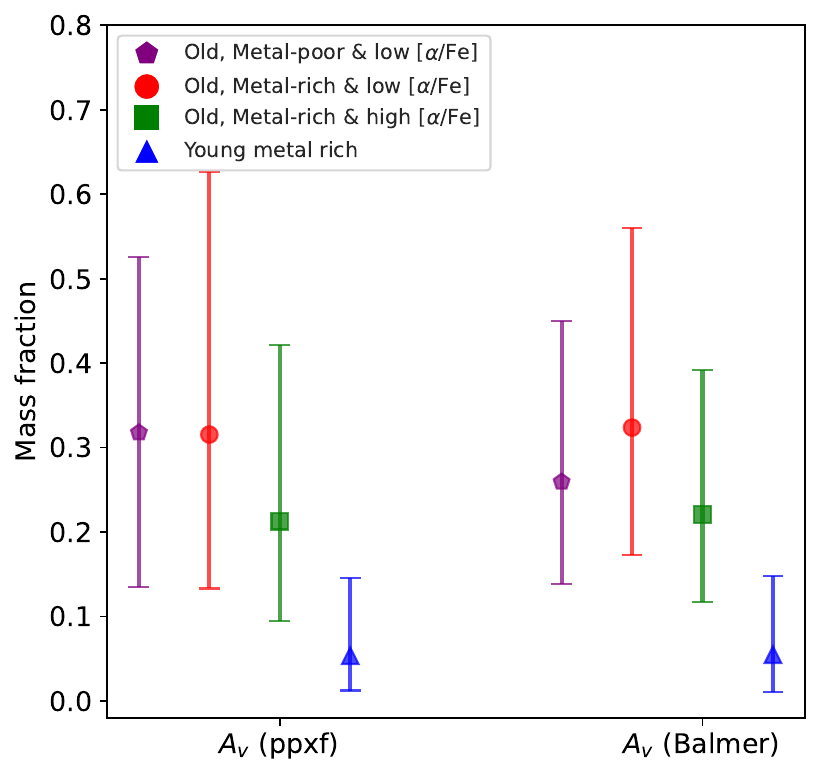}
    \includegraphics[width=0.7\columnwidth]{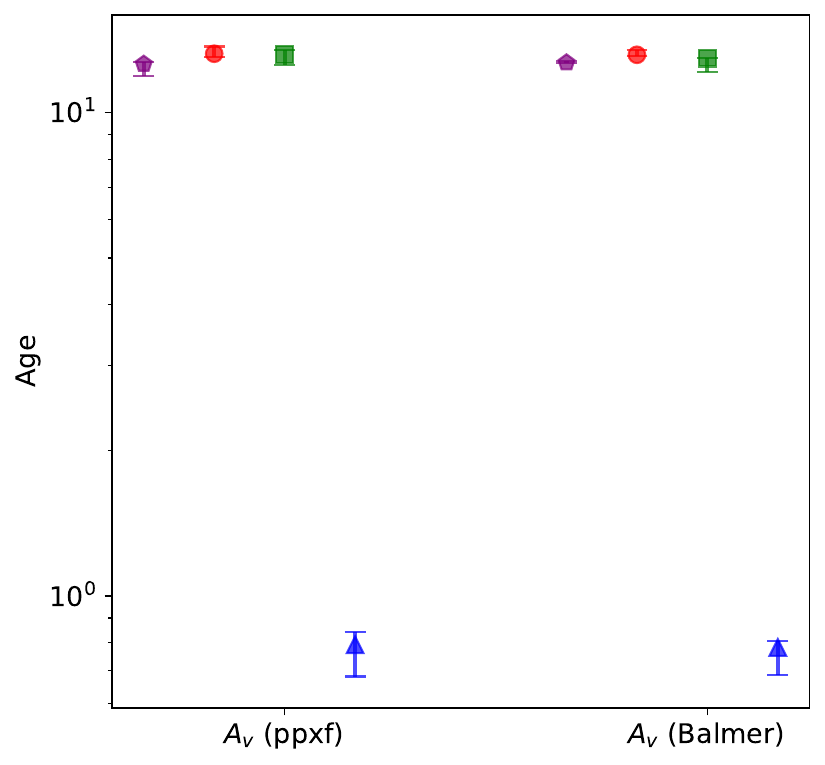}
    \includegraphics[width=0.7\columnwidth]{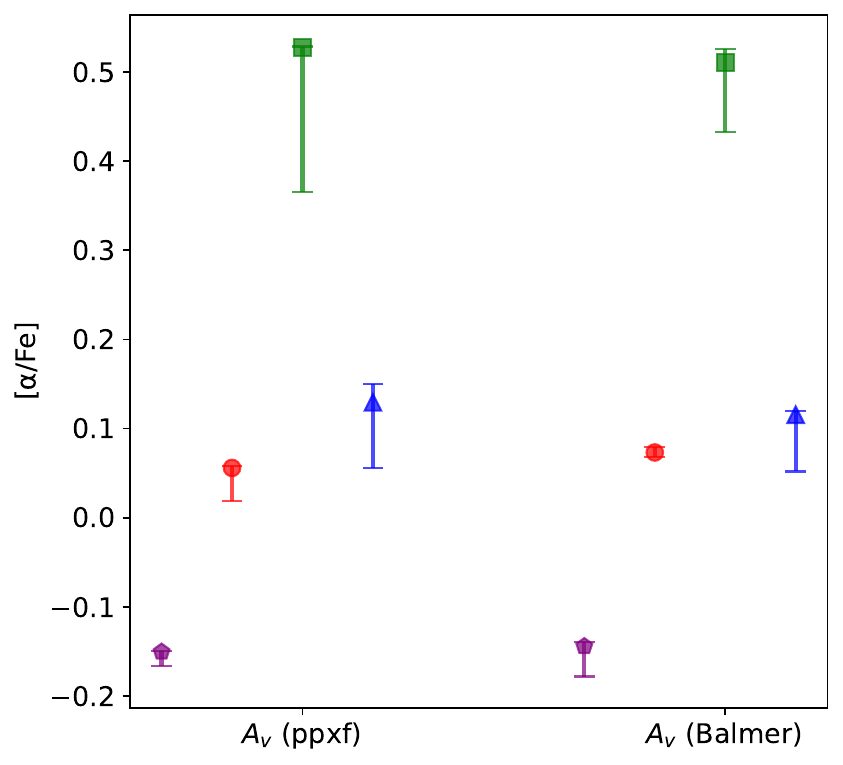}
    \includegraphics[width=0.7\columnwidth]{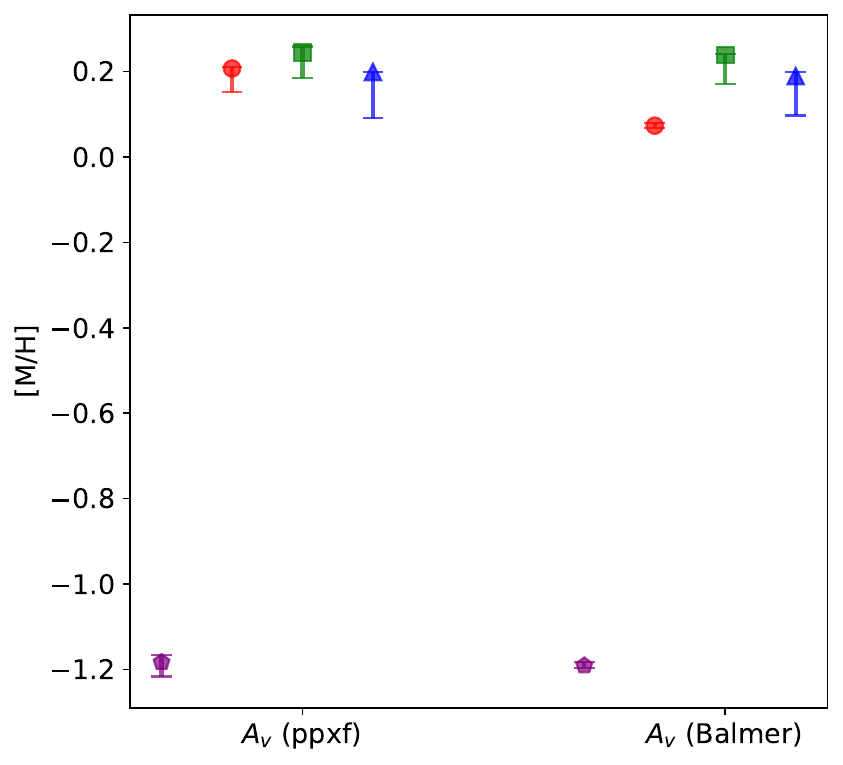}

    \caption{Mean properties (mass-fraction [top left], age [top right], [$\alpha$/Fe] [bottom left], [M/H] [bottom right]) of the stellar populations (old, metal-poor and low [$\alpha$/Fe] -- purple; old, metal-rich and low [$\alpha$/Fe] -- red; old, metal-rich and high [$\alpha$/Fe] -- green; younger metal-rich -- blue) for an example bin of ESO 544-27, separately for $A_{v}$(ppxf) and $A_{v}$(Balmer). }
    \label{Fig:Error Bar bin zero}
\end{figure*}

    We check the effect of dust attenuation on the stellar population parameters derived from full spectral fitting with pPXF. As discussed in Section~\ref{sect: analysis}, we have utilised the A$_v$ estimated by pPXF to correct each spectra for dust attenuation. However, stellar dust attenuation estimation from stellar continuum fitting is complicated and depends on the underlying stellar population and the extinction law being utilised. When both H$\alpha$ and H$\beta$ emission line fluxes are measured, we can derive the nebular dust extinction at the location of the stellar extinction (from the same spatially binned spectra) using the Balmer decrement. While the nebular extinction serves as a useful identifier of the presence and spatial distribution of dust in the galaxy, it has been found to not directly correlate with stellar extinction (see Figure 12 and associated text in \citealt{2017Viane}). Nevertheless, the nebular extinction can be used for estimating the stellar extinction in the same spatial bin from the relation by \citet{Calzetti94}. We thus utilise the Balmer decrement to dust-correct the spectrum of a single spatial bin to in ESO~544-27 assuming the \citet{Calzetti94} reddening law and compare the derived stellar population properties with that derived from the pPXF dust attenuation corrected spectra for this spatial bin. 
    
    In Figure~\ref{Fig:Error Bar bin zero}, we compare the mass fraction, mean age, mean [M/H] and mean [$\alpha$/Fe] for the different stellar populations, and present the differences introduced by utilizing the Balmer decrement compared to the pPXF estimated A$_V$. The old metal-poor low-[$\alpha$/Fe] population  shows a lower mass-fraction, albeit within error, for the A$_V$ estimated with Balmer decrement. The mean properties of the stellar populations remain largely unaffected by including dust attenuation correction from the Balmer decrement in the full spectral fitting. Only the mean [M/H] of the old metal-rich, low [$\rm\alpha$/Fe] stellar population shows a slight decrease of $\sim$0.1 dex. Since the dust attenuation correction from the Balmer decrement does not affect the derived mean properties significantly, we have carried out the stellar population analysis as described in Section~\ref{sect: analysis}. 


\bsp	
\label{lastpage}
\end{document}